\documentclass[transmag]{IEEEtran}
\usepackage{latexsym}
\usepackage{graphicx}
\usepackage{amsfonts,amssymb,amsmath}
\usepackage{hyperref}
\usepackage{pifont}
\newcommand{\cmark}{\ding{51}}%
\newcommand{\xmark}{\ding{55}}%
\usepackage{dirtytalk}
\usepackage{textcomp}
\usepackage{multirow}
\usepackage{adjustbox}
\usepackage{array}
\def\BibTeX{{\rm B\kern-.05em{\sc i\kern-.025em b}\kern-.08em T\kern-.1667em\lower.7ex\hbox{E}\kern-.125emX}}
\newcolumntype{P}[1]{>{\centering\arraybackslash}p{#1}}
\begin{document}

\title{Artificial Intelligence for UAV-enabled Wireless Networks: A Survey}
\author{Mohamed-Amine Lahmeri, \IEEEmembership{Student Member, IEEE}, Mustafa A.Kishk,  \IEEEmembership{Member, IEEE},\\ and Mohamed-Slim Alouini, \IEEEmembership{Fellow, IEEE}
	\thanks{Mohamed-Amine Lahmeri, Mustafa A. Kishk, and Mohamed-Slim Alouini are with King Abdullah University of Science and Technology (KAUST), Thuwal 23955-6900, Saudi Arabia (e-mail:mohamed.lahmeri@kaust.edu.sa ; mustafa.kishk@kaust.edu.sa; slim.alouini@kaust.edu.sa).} 
}

\IEEEtitleabstractindextext{\begin{abstract}Unmanned aerial vehicles (UAVs) are considered as one of the promising technologies for the next-generation wireless communication networks. Their mobility and their ability to establish line of sight (LOS) links with the users made them key solutions for many potential applications. In the same vein, artificial intelligence (AI) is growing rapidly nowadays and has been very successful, particularly due to the massive amount of the available data. As a result, a significant part of the research community has started to integrate intelligence  at the core of UAVs networks by applying AI algorithms in solving several problems in relation to drones. In this article, we provide a comprehensive overview of some potential applications of AI in UAV-based networks. We also highlight  the limits of the existing works and  outline some potential future applications of AI for UAV networks.
	\end{abstract}

}

\maketitle

\section{INTRODUCTION}

\IEEEPARstart{U}{nmanned} aerial vehicles, known as UAVs, attracted a lot of research interest in the last decades due to many inherent attributes such as their mobility, their easy deployment, and their ability to establish line of sight (LOS) links with the users~\cite{Survey1,Survey2,Survey3,Previouswork4}. In general, UAVs can be classified into two main types, namely fixed-wing and rotary-wing UAVs. Each type of UAV is adapted to a specific type of application. For example, fixed-wing UAVs are more appropriate for the type of missions where stationarity is not required, e.g. military applications such as attack and surveillance. However, rotary wing UAVs have more complex aerodynamics. They also have the ability to remain stationary at a given location, but they cannot carry out long-range missions.   For example, rotary-wing UAVs are better suited to provide temporary wireless coverage to ground users.\par 
The involvement of many industries in the manufacture of UAVs has helped to reduce their cost on the markets, making the use of a UAV network no longer a dream or a futuristic idea. In fact, they have been used in many scenarios such as providing wireless connectivity, weather forecasting, disaster management, farming, delivery and traffic control~\cite{UAVapp1,UAVapp2,Imaging13,Imaging16,Imaging15,ML1}.\par 
A number of limitations related to the use of UAVs can be raised, such as their vulnerability to severe weather conditions, the need for LOS link to avoid a risky loss of control. Most importantly, the constrained battery and low computational capabilities of drones are considered as their main limitations~\cite{UAVlimit,Previouswork8,Previouswork9,Previouswork10,Previouswork11,Previouswork12}. In fact, most commercially available UAVs struggle to hover for more than two hours and must always return to base to recharge their batteries. Added to that, the fact that complex algorithms cannot run onboard due to the limited computational capacity, consequently, classical solutions and algorithms does not necessarily fit any UAV-related problem.\par 
In another context, machine learning (ML) has emerged in the last years as a sub-field of AI. Moreover, its use has become prevalent in the scientific research offering a new style usually referred to as the black-box technique where you only care about inputs and outputs. Furthermore, the huge amount of data available nowadays and the existence of the high performance computing (HPC) and good GPU's helped ML to see the light. As a result, ML is being actively used today in many fields, perhaps more than one would expect.\par 
We can also notice the emergence of several sub-fields from AI such as deep learning (DL),  reinforcement learning (RL), and federated learning (FL), each for a specific type of problems. For example, DL is a branch of AI that uses layers of artificial perceptrons to imitate the human mind thinking. It is massively used in speech recognition, computer vision, and natural language processing. RL is another branch of AI that appeared around 1979~\cite{RLbook} wherein an agent learns the way of making good action in order to achieve maximum rewards. The learning process is achieved by exploitation and exploration of the different available states. RL is an active field of ML that evolved and matured very quickly. 
Unlike DL, RL is massively used in robotics for path planning and learning the way to do complex tasks. This does not mean that it is limited only to robotics, it is also used in many other decision-making problems that consider a goal-oriented agent that  is interacting with a specific environment. 
Another new field of ML is FL, which was proposed by Google in 2016 and designed to support network systems with decentralized data. FL is considered as an ML setting with the objective of training a highly centralized model on devices sharing decentralized data without the need of sending the data to a local shared unit. In other words, it is used to run ML algorithms with decentralized data architecture. This task is performed in a secure manner and when it comes to UAV-based networks, the use of FL is indeed a hot topic.\par 
In short, AI is one of the trending areas that brings intelligence to machines and makes them able to perform tasks even better than a human can do. It is  believed that combining the advantages of using AI within UAVs networks is a challenging and interesting idea at the same time. 
In the same vein, UAV-based applications can be improved by integrating AI at the core of UAV networks. For instance, as mentioned earlier, UAV batteries are limited, therefore AI can play an important role in resource management for UAVs with the aim of maximizing energy efficiency~\cite{RL9}. Moreover, the design of UAVs trajectory and deployment are also subject to AI improvement by equipping the UAV with the ability to   design its trajectory automatically.  
Imaging also can be improved for UAVs by applying the existing state of art related to computer vision for UAVs imaging. A wide range of applications can be improved in this context such as surveillance, traffic management, and landing site detection.\par
To conclude, UAV-based networks performance can be highly improved with the integration of AI algorithms in order to automate complex tasks and enhance the overall system level of intelligence.

\subsection{Previous Surveys and Tutorial Works:}\par
With the vast amount of published work linking AI to UAV wireless networks, several tutorials and surveys have attempted to  summarize the existing literature.
The authors in~\cite{Previouswork4} provided a tutorial for UAV-based wireless communication systems by covering potential applications, challenges and describing the open problems in the field. However, the aforementioned work does not focus on the ML aspect for UAVs.
Many other surveys and tutorials do not focus on AI techniques, for instance, a motion planning for UAVs guide was presented in~\cite{Previouswork5} and a survey for UAV traffic monitoring is provided in~\cite{Previouswork6}.\par
In addition to the work mentioned above, there exist other tutorials and surveys that are oriented towards the application of ML tools in wireless communication networks. For example, a tutorial on artificial neural networks (ANNs) for wireless networks is proposed in~\cite{Previouswork5}. In the same context, a review for RL and DL techniques for UAVs summarized some of the works done in this area~\cite{Previouswork2,Previouswork0}. However, all the works mentioned above does not consider AI techniques specifically for UAVs applications.\par  
Nevertheless, one work can be considered as close to this work, namely the ML study for UAV communication presented in~\cite{Previouswork3}. However, there are still major differences between the two works. First of all, the structure of this survey is different from that of~\cite{Previouswork3} since we review each AI subfield separately and delve into the different UAV-based applications of each AI area, whereas in~\cite{Previouswork3}, the survey is based on the application type. We believe that this structure not only allows us to explore each area of AI in depth, but also to provide a comprehensive overview of each AI area by presenting the most commonly used algorithms in UAV-related problems. Therefore, this survey is aimed at all kinds of readers, including those who have no knowledge of AI and also for researchers who have started to take an interest in AI from a wireless communication perspective. In addition, the work in~\cite{Previouswork3} did not cover FL, which, in our opinion, is a key technique for bringing intelligence to the edge of UAV networks in a decentralized and secure way. Given the significant importance of FL, particularly for 6G networks, we covered this area in depth by reviewing the recently published work.
We also maintained a unique approach in each AI domain and focused on the most recent publications. For example, one of the most prominent applications covered in this survey is the fusion of RL with the use of intelligent reflective surface (IRS)-equipped UAVs to support millimeter wave bands which was not covered in~\cite{Previouswork3}.

\subsection{Contributions}\par In this article, we provide a holistic overview of the state of the art research in the area of AI-enabled UAV networks. We also discuss some limitations of the existing research works and outline some potential ideas that could be addressed in the near future. We also study the implementation of intelligence at the edge of UAV networks by reporting some of the works done in FL for UAV-based networks. Furthermore, we provide a comprehensive introduction to each AI area studied in this work so that readers with different backgrounds have the ability to understand this article.\par
This survey is organized as follows:\\ 
\begin{itemize}
	\item In  Sec.~\ref{sec:SuperAndUnsuper}, we start by reporting the supervised and unsupervised ML works designed for UAV-based networks. A brief overview will cover these two different areas of ML and some typical algorithms and neural network (NN) architectures be provided for the reader's  convenience.
	\item In Sec.~\ref{sec:RL}, we go over the works relating RL with UAV-related problems. We start by a quick overview of RL basics and  present a classic example for RL path planning in order to understand the basic concepts of this area.
	\item In Sec.~\ref{sec:FL}, after introducing FL, we outline the key research directions that enable installing intelligence at the edge of a UAV network by reporting some of the works relating FL to UAV-related problems. 
\end{itemize}

All the above-mentioned sections are concluded with a discussion presenting the limitations of the current research work and highlighting some possible future works that could be established. 
 \begin{figure*}
	\centering
	\includegraphics[width=0.9\textwidth]{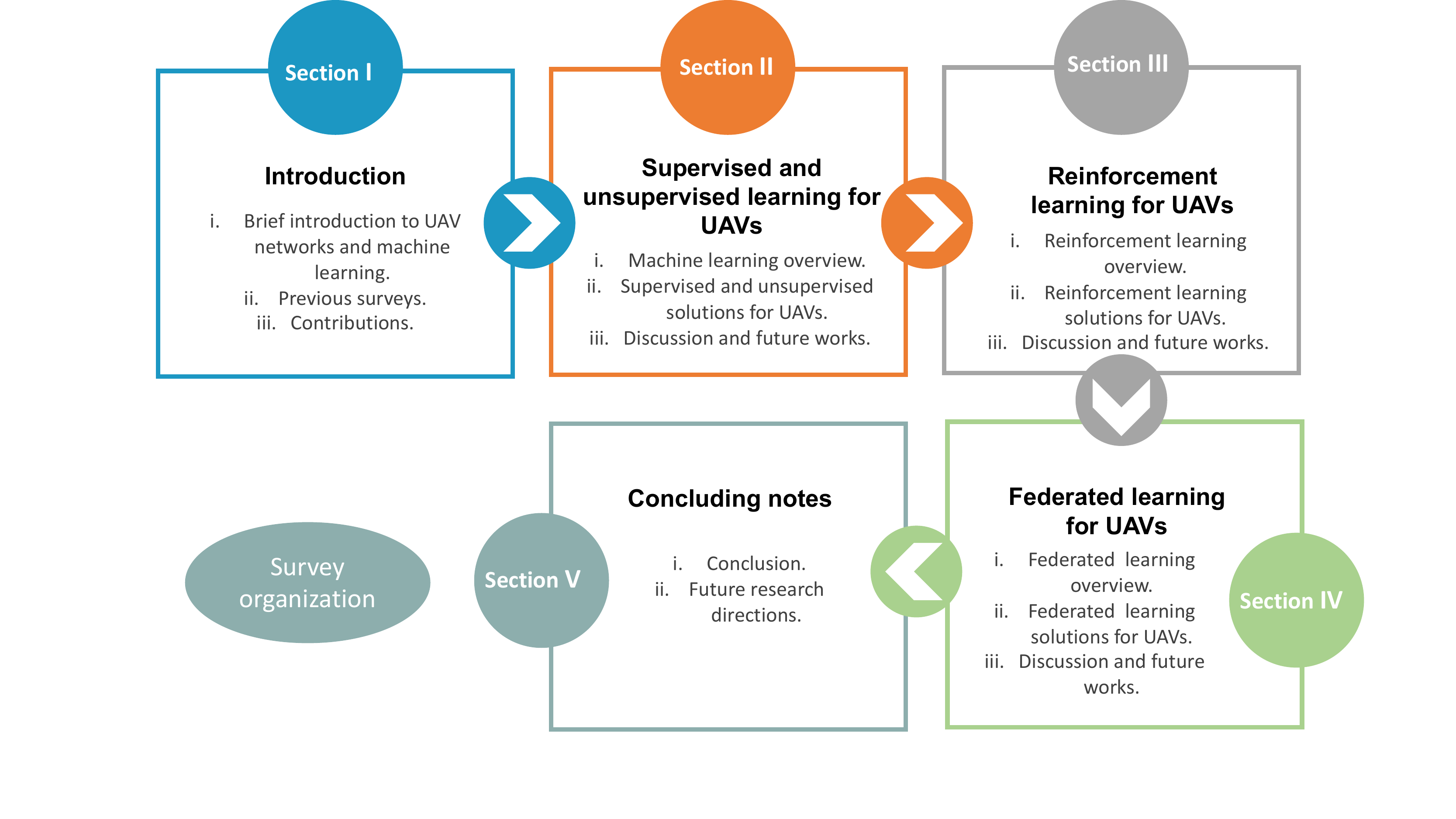}
	\caption{Survey organization.}
	\label{fig:NN-architectures}
\end{figure*}

\section{Supervised and Unsupervised machine learning for UAVs} \label{sec:SuperAndUnsuper}
\begin{figure}
	\centering
	\centerline{\includegraphics[width=3.5in]{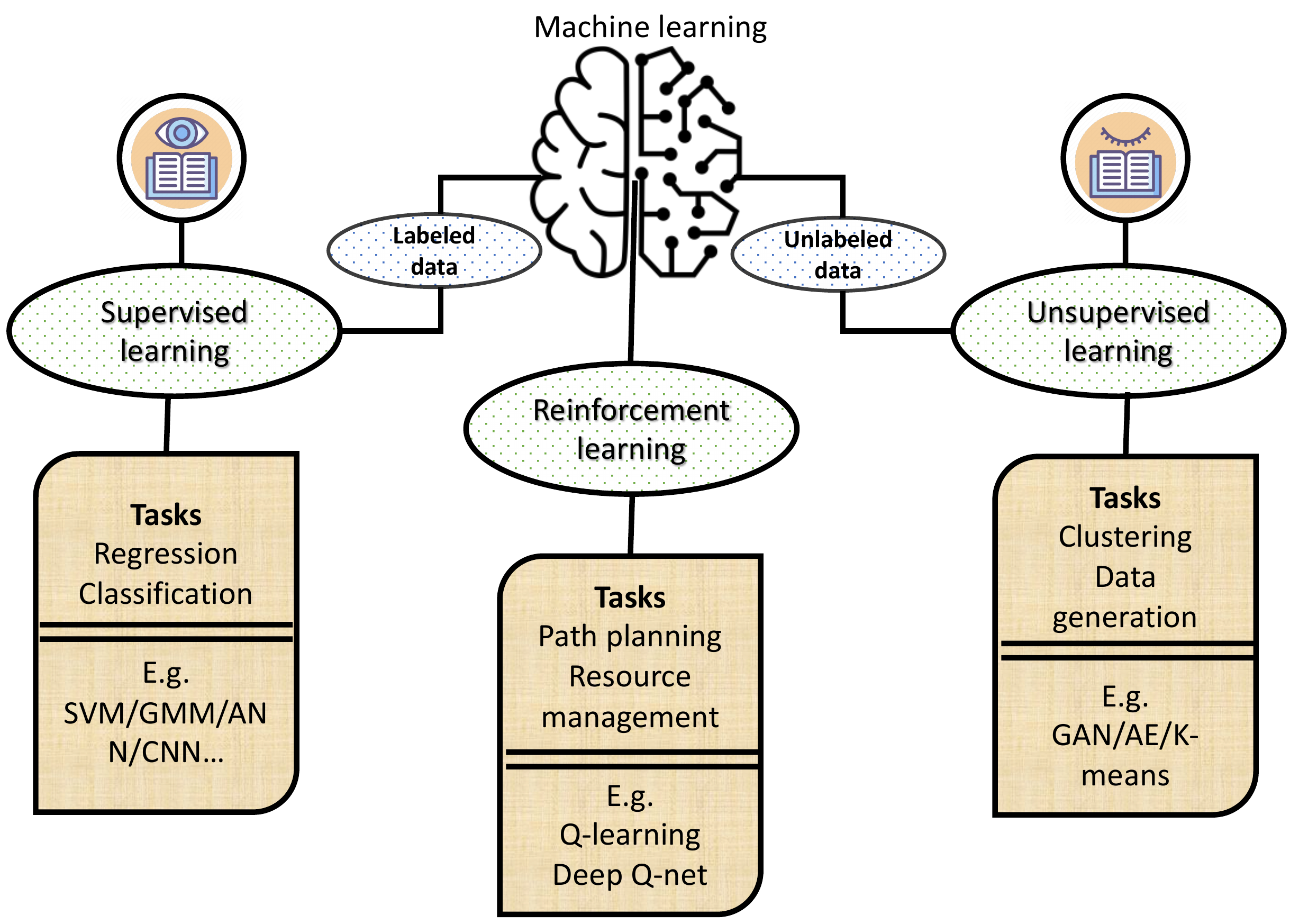}}
	\caption{Machine learning overview.}
	\label{fig:machine-learning-overview}
\end{figure}
ML is a recent buzzword related to AI. In short, it is the subset of AI that enables a computer to execute tasks accurately based on the experience gained by learning from some previous examples. In fact, ML has been very successful over the last decade because of the large available amount of data and today's powerful computers. This is why research is now oriented towards applying ML in UAV-based problems. .\par
The areas of ML can be divided into different categories of problems, for instance, it might be divided as shown in Fig.~\ref{fig:machine-learning-overview}  to supervised learning problems, unsupervised learning problems, and RL-based problems. In what follows, we" distinguish between the supervised and the unsupervised learning areas to avoid confusion later. 

\subsection{Supervised Learning Overview }
\label{Sec:SupervisedML}
In supervised learning, the data provided is labeled, in other words, we provide for each data entry the ground-truth value so that the algorithm uses these values to learn how to make a decision for a new unlabeled entry. For example, predicting a UAV price from its characteristics. In this example, you need to provide the algorithm with a set of training data that contains each UAV characteristics and its associated label (the price). The dataset is usually divided into a  training set and a test set. The training set is used to learn the relationship between the input and the output and the test set is used to validate the model by measuring its accuracy. The supervised problems are often divided into either regression problems or classification problems. Regression problems provide continuous output values (eg. predicting a price). However, classification problems provide discrete values indicating to which class the input belongs to (eg. classify benign or malignant cancer disease). In what follows, we present the most well-known ML algorithms for supervised and unsupervised learning. We also focus on the algorithms that are used to solve the UAV-related problems reported in this survey. \\ 
\begin{figure*}
	\centering
	\includegraphics[width=5in]{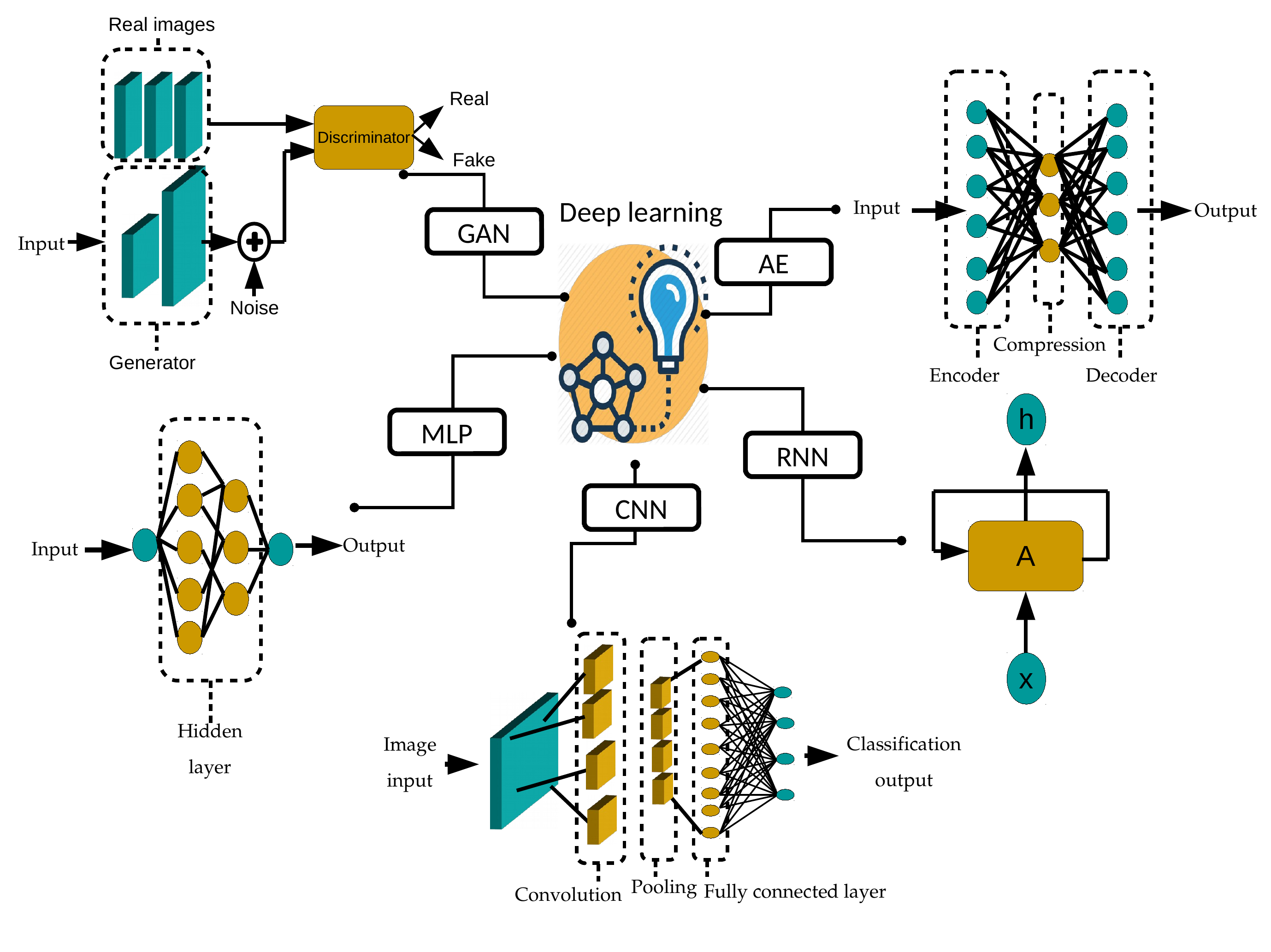}
	\caption{Neural network architectures.}
	\label{fig:NN-architectures}
\end{figure*}

\textbf{Some supervised algorithms and NN architectures:}
\begin{enumerate}
	\item \textbf{Combined classification and regression algorithms}\\
	There are several supervised algorithms that can be used either for classification or regression. For instance Support Vector Machine (SVM) can do both the tasks, decision trees also can be formulated to solve regression or classification depending on the use case. 
	\item \textbf{Regression algorithms}\\
	There exists algorithms that perform pure regression task by predicting continuous value output. For instance, we can mention two classical algorithms in ML which are  the linear regression and the logistic regression.
	\item \textbf{Classification algorithms}\\
	It makes sense to talk about pure classifiers in ML. Although it is mentioned in some references that Naive Bayes classifier with "some modification" can be used for regression, we present it as a pure classifier example since it was derived initially for classification based on the probabilistic Bayes theorem. 
	\item \textbf{Multi Layer Perceptron (MLP)}\\
	To imitate the biological human neural networks, ANNs are mathematically formulated for ML. ANNs are built with a number of partially-connected nodes denoted by perceptrons and grouped into different layers. Each perceptron is responsible for processing information from its input and delivering an output. As shown in Fig.~\ref{fig:NN-architectures}, MLP is the simplest form of an ANN that consists of one input layer, one or more hidden layers, and an output layer where a classification or regression task is performed.
	\item \textbf{Convolutional neural networks (CNNs)}\\
	CNN is another type of ANN designed initially for computer vision tasks. A CNN usually takes an image as an input, assigns learnable weights and biases that are updated according to a specific algorithm. The CNN architecture is characterized by the convolutional layers which extract high-level features from the image that will be used later. Technical details such as activation functions, pooling layers, and padding operation are beyond the scope of this survey. Fig.~\ref{fig:NN-architectures} shows a typical CNN architecture where feature extraction is performed in the first convolutional layers and classification is performed via a fully-connected layer. 
	\item \textbf{Recurrent neural networks (RNNs)}\\
	When the data is sequential in nature, RNNs take place to solve the problem. For the sake of example, we can mention a text speech, a video or a sound recording. 
	RNNs are widely used in natural language processing (NLP), in speech recognition, and for generating image description automatically.
	The RNN architecture is similar to a regular neural network, only it contains a loop that allows the model to carry forward results from previous neurons.  RNN in its simplest form is composed of an output containing the prediction and which is denoted by h in Fig.~\ref{fig:NN-architectures} and a hidden state that represents the short term memory of the system. 
	
\end{enumerate}

\subsection{Unsupervised Learning Overview }
Unlike supervised learning, the unsupervised learning does not use labeled data, instead, it looks for some underlying structure or hidden pattern in the data and reveals it. For instance, clustering the data, reducing data dimensionality, and data generation are considered as typical tasks for unsupervised learning. In what follows, we present some classical unsupervised algorithms.\\
\textbf{Unsupervised algorithms and NN architectures:
}

\begin{enumerate} 
	\item	\textbf{$K$-means}\\
	$K$-means is a very popular algorithm for clustering in ML. It takes a number of clusters $K$ as an input and  allocates every data point to the nearest cluster, while keeping the centroids as small as possible.
	\item \textbf{Gaussian Mixture Modeling
		(GMM)}\\
	GMM is another clustering algorithm in ML, but unlike the $K$-means algorithm, GMM is a probabilistic model. As its name indicates, the clusters are derived from a Gaussian distribution and the association is soft. In other words,  every data point have a probability of association to every cluster center, however, in the $K$-means algorithm, we have hard association policy. 
	\item \textbf{Autoencoders (AEs)
	}\\
	AE is a type of neural network used to learn a representation of the data and hence encode it. This technique is often used for dimensionality reduction. Surprisingly, the architecture of an AE is extremely simple as the Fig.~\ref{fig:NN-architectures} shows. It is usually formed by an input layer and a hidden layer called "bottleneck" which forces a compressed knowledge representation of the original input. 
	\item \textbf{Generative adversarial networks (GANs)
	}\\
	GANs are algorithmic architectures that use two neural networks in order to generate new, synthetic instances of data that can pass for real data. They are used widely in image generation, video generation, and voice generation.

\end{enumerate}

\subsection{Supervised and Unsupervised Solutions for UAVs-based Problems}
\subsubsection{The positioning and deployment of UAVs}
Authors in~\cite{ML1} investigated the optimal deployment of aerial base stations  to offload terrestrial base stations by minimizing the power consumption of the drones. The provided solution  is considered as ML-assisted due to the fact that UAVs are not required to continuously change their positions, instead, they are placed temporarily  by predicting the congestion in the wireless network. The wireless traffic is predicted based on the GMM which is a probabilistic model that belongs to the set of unsupervised ML defined previously. It assumes that the data distribution can be modeled by the Gaussian distribution. First, a $K$-means algorithm divides the users into $K$ clusters and then a weighted expectation maximization algorithm is performed on the $K$ clusters in order to find the optimal parameters for the GMM model.
The next step is to deduce the optimal deployment by formulating a power minimization problem for the UAVs. 
The numerical results show that the ML-assisted approach outperforms the classical solution by reducing the mobility and the power needed for downlink purposes. Although the work done is of great importance by combining ML with optimization techniques, using a $K$-means algorithm to classify the users brings the question of how to choose manually the value of the number of clusters $K$ and also how to initialize the cluster center positions.\par
In the same context, the authors in~\cite{ML6} investigate an optimal placement of UAVs acting as aerial base stations by building a structured radio map. Due to the nature of the complex terrain and the difficulty of exploiting such radio map, the authors proposed a joint clustering and regressing problem using a maximum likelihood approach that is formulated based on the $K$-segment ray tracing model. ML is also used in predicting the channel in order to reconstruct the radio map.\par
The optimal deployment of aerial base stations using UAVs is also studied in~\cite{ML24}. ML techniques are used to predict download traffic using weighted expectation maximization. This ML technique has been compared in terms of performance to a baseline expectation maximization algorithm and the $K$-means algorithm. In addition, contract theory was used to ensure that the downlink demand is satisfied by selecting the appropriate UAV for each hotspot.\par
In~\cite{ML7} the communication efficiency between a UAV and a base station is improved by predicting the location of the UAV given its past locations. In fact, while offloading a terrestrial base station, a UAV can be subject to wind perturbation which will result in a certain degree of offset and hence a loss in capacity. To solve this issue, the authors propose a RNN-assisted framework where the next elevation and horizontal angles of the UAV with reference to the base station are predicted using the past angles. This method leads to predicting the specific location for a high-speed movement UAV. The authors keep tuning the RNN parameters such as the number of hidden nodes and the number of hidden layers and then study their impact on the prediction accuracy. Numerical results have shown that a high accuracy could be achieved for a 4 layer RNN with 16 hidden nodes.   

Further alternatives for UAV path planning have been proposed in the open literature. In~\cite{ML22} an unsupervised solution has been proposed to enable motion prediction for a group of heterogeneous flying UAVs.  In addition to predicting the future locations of the UAVs, the algorithm was designed to classify the network nodes based on their motion properties. In addition, the authors in~\cite{ML23} used ANN to predict the optimal location of UAVs that are used as relays in a Flying Ad Hoc Network (FANET) setup. 

\subsubsection{Channel estimation}
 One could think about how ML can improve the well-established empirical models used to estimate and model the complex UAV-to-UAV and ground-to-UAV channels. In this context, the work in~\cite{ML2} investigated the prediction of the UAV-to-UAV path loss. The predictions generated by the $K$ Nearest Neighbors (KNN) and the Random Forest algorithms are compared to empirical results. The path loss is predicted starting from several parameters such as the propagation distance, transmitter altitude, receiver altitude and elevation angle. The comparison of  the results with the data generated by the ray-tracing software shows that ML performs well in these prediction tasks.\par 
As millimeter wave bands are being exploited for next-generation cellular systems to improve the communication capacity, the authors in~\cite{other4} use a generative neural network to predict the channel state between a UAV and two types of terrestrial base stations: (i) terrestrial street-level base stations and (ii) aerial roof-mounted base stations.  A first neural network classifies the link type (LOS/NLOS/outage) and feeds this information to a second neural network to generate the different channel parameters.\par
Supervised ML was used in other research works to predict channel quality between a UAV and ground nodes. For instance, ANNs were used to predict the path loss in~\cite{ML3,channel1,channel4,channel5}. The authors in~\cite{ML3} use ANNs to predict the signal strength of the UAV and estimate the channel propagation.
A shallow ANN is proposed to analyze the effect of several natural phenomena on the signal such as : diffraction, reflection, and scattering. The input layer is composed of parameters like the distance to the UAV, altitude, frequency, and path loss. This exciting work may be impeded by the large  processing time of the data by the ANN which raises the question of whether this solution is adequate for real-time applications or not. In the same context, in~\cite{channel5}  the signal strength between the UAV and ground nodes is estimated using ANN. The authors considered an urban environment where the signal strength data was used as an input to feed an ANN. Using this data, the channel parameters were estimated accurately.  
In addition to ANN, other supervised ML algorithms were used in~\cite{channel2} to predict the received signal strength received at a flying unit from a cellular base station.\par  In contrast, in~\cite{channel3} unsupervised learning was used for a 3D channel modeling between UAVs and ground mobile users. The popular $K$-means algorithm was used in this work to classify the links into LOS and NLOS. Aside from ANN, the SVM algorithm  can be used for regression in addition to classification. The   work in~\cite{ML8} proposes a method  for path loss prediction in urban outdoor environment using support vector regression algorithm and compares the obtained results with the empirical ones.  

\subsubsection{UAV detection}
Since UAVs are used extensively not only by the military but also by civilians, several applications need to be supervised by the authorities because UAVs could be used for espionage or even as a lethal weapon. Therefore, the detection and tracking of UAVs is very important to limit these dangers. In this field, numerous research projects have proposed different solutions, which we divide into image-based solutions and sound-based solutions.\par 
Starting with sound-based detection problems,
the authors in~\cite{ML4} present a real-time UAV detection system based on analyzing the sound data coming from the drone. For this purpose, two ML methods have been applied and compared in terms of accuracy.
The first step consists in detecting potential UAV existing by analyzing the frequency and then check whether the sound exceeds a predefined threshold for drones. The first ML method used is Plotted Image Machine Learning (PIL). This method uses the visualized Fast Fourier Transform (FFT) graph generated from the data sound to compare average image similarity with a reference FFT target. 
The second method is based on the KNN algorithm applied to the FFT and measures the average distance with the target. 
The simulation results show that the PIL method outperforms the KNN methods and succeeded to provide good results. At this level, we point out that even if the visual drone detection can be limited by the quality and the resolution of the input image, sound data also can be highly affected by ambient noise in real applications. It also is not obvious if all the UAVs will have the same FFT profile used as predefined target in the problem. Moreover, KNN algorithm is a simple and straightforward algorithm in ML and hence trying more sophisticated algorithms will be beneficial for the problem.
In another line of research, the authors in~\cite{ML10} used a spiral microphone array to detect and track the position of a flying UAV.  Several spectrograms and filters are applied to the input sound to perform feature detection before feeding the concurrent neural networks architecture to perform the task. The scientific literature is still rich with regard to the recognition of the acoustic signature of UAVs~\cite{ML13,ML14,ML15}. Other research works used different features to detect the UAV presence such as the WiFi traffic~\cite{ML16} or the UAV radio frequency signal~\cite{ML17,ML18,ML19,ML20}.\par Another approach to address UAV detection problem is through images instead of sound. In this context, different deep ML architectures were compared at the visual UAV detection task using a Pan-Tilt-Zoom camera in~\cite{ML11}. Moreover, a recent work in~\cite{ML12} reviewed some of the computer vision techniques used in UAV detection problems. 

Finally, we highlight the fact that making a hybrid system that uses image, sound, and radio UAV transmission signals at the same time will be a very interesting futuristic idea. In this context, we refer interested readers to the recent work in~\cite{ML4}.   

\subsubsection{Imaging for UAVs}

\begin{table*}[]\caption{ Imaging for UAVs. }
	\centering
	
		\label{Tab:table1}
		\begin{tabular}{|c|c|c|c|}
			
			\hline
			Reference & Model & Application & Date of publication \\ \hline
			\cite{Imaging3}          &Faster R-CNN      & Car detection      & Late 2017      \\ \hline
			\cite{Imaging4} &Nazr-CNN        &Damage detection          &Late 2017   \\ \hline
			
			\cite{Imaging1}         & CNN+SVM=CSVM   &General object detection         & 2018     \\ \hline
			\cite{Imaging14}                 &modified region-based CNN
			&Electrical equipment defect detection 
			&2018   \\ \hline
			
			\cite{Imaging5}          & Faster R-CNN+ Region proposal network(RPN)
			&Pedestrians detection &2018       \\ \hline
			\cite{Imaging6}          &Semantic segmentation+CNN     &UAV geolocalization      &2018       \\ \hline
			\cite{Imaging7}        & CNN   &Car detection         &2018 \\ \hline

			\cite{Imaging11}                 &CNN          &Building crack detection   &2018   \\ \hline
			\cite{Imaging8}   &Faster R-CNN+Yolov3+RetinaNet
			&Tree detection                &2019\\ \hline
			\cite{Imaging12}                 &CNN+Digital Surface Model(DSM)          &Surface classification   &2019   \\ \hline
			\cite{Imaging2}         & CNN with semi-supervised learning    & Agricultural detection (soybean leaf and herbivorous pest)
			& 2019     \\ \hline
			\cite{Imaging13}                 &CNN          &Rice-grain estimation   &2019   \\ \hline
			
			\cite{Imaging15}                 &Yolov3
			&Weed location 
			&2020   \\ \hline
			\cite{Imaging16}                 &CNN
			&Single tree detection
			&2020  \\ \hline
			\cite{Imaging17}                 &Yolov2
			&Green mangoes detection
			&2020  \\ \hline
			\cite{Imaging10}                  &Faster R-CNN          &Maize Tassels detection   &2020  \\ \hline
			\cite{Imaging9}                 &CNN          &Counting and locating citrus trees  &2020   \\ \hline
		\end{tabular}

	\label{table:1}
\end{table*}

Although computer vision is beyond the scope of this survey, several research works relating imaging to UAVs exist in the literature, for example, authors in [11] investigates the detection of a forced (emergency) safe landing site. The detection is converted into a classification problem where two known classifiers ( GMM and SVM ) are tested. The classifier converts the real map into a safe or unsafe grid map. A filter is applied later to remove unsafe spots and keep the potential landing sites. The main reason why these types of problems are not considered in this study is that they can be treated as pure computer vision problems, and the application to UAVs does not change the nature of the task, except that the images are taken from a given altitude. In other words, the same techniques are applied to UAVs imaging, such as  CNN, feature extractors, and edge detectors.
For the reader convenience, we summarize a number of the recent works in Table~\ref{Tab:table1}. We also refer the readers interested in more UAV imaging problems to the Table.1 in~\cite{DLreview}. 

\subsection{Discussions and Future Works}

\subsubsection{Practical issues of ML implementation}
The application of ML techniques in UAV-based networks can be hampered by the limited computing capacity onboard. In fact, most commercially available UAVs are not equipped with the sophisticated processors needed to execute heavy ML algorithms. 
Even if you intend to equip a drone with a powerful CPU and GPU, you must first take into account its weight and power consumption. As a result, the same problem will persist due to the power constraints of UAVs. 
One solution to this problem is to use the cloud to train models and make inferences at the UAV level. However, this solution will increase the communication costs, which in turn will bring us back to the energy constraint problem, because the UAV will have to communicate back and forth with the cloud. 
Therefore, another interesting solution to the problem is to run the ML onboard, but this time adjusting the ML algorithms to the UAV\'s limited capacity. This technique leads us to a new field usually referred to as on-device learning dedicated to constrained devices. A recently published study~\cite{ML0} has examined device learning by addressing lightweight ML algorithms and discussing the different ML and DL algorithms in terms of complexity and resource consumption. 
We propose a final solution to address the execution of ML onboard, namely FL. It consists of executing ML in a decentralized way by sending model updates over networks instead of sharing raw data. We intend to cover and discuss this technique at  Sec.\ref{sec:FL} of this survey.\par
In addition to the hardware and software limitations of UAVs mentioned above, the practical use of ML in UAV networks still faces other significant hurdles related to existing rules and regulations. Although research is aimed at partially or even fully autonomous UAV applications, most existing regulations do not allow such operations in real life. For example, the U.S. Federal Aviation Administration (FAA), in its latest regulation published in December 2020, did not specify a single point concerning autonomous UAVs. Instead, it focuses on regulations dedicated to the human operators who control a drone. That being said, it is important to mention that there is still hope for autonomous UAVs to see the light of day. In fact, unlike the FAA, the European Aviation Safety Agency (EASA), in its latest regulation published in December 2020, admits the existence of autonomous UAV operations by including them and classifying them in different classes according to the risk level of the application. This will certainly offer new opportunities for innovative UAV solutions based on ML and AI in general. In conclusion, it is important to harmonize and unify the UAV regulations around the world, as this will motivate future research in this area.
\subsubsection{Literature discussions}Although we have tried to criticize objectively some of the work we have covered previously, in this section we intend to present our thoughts on the use of ML in wireless communication problems.\par 
Firstly, it can be noted that, in the literature, ML tools are often used to solve problems that could be solved in a simpler and more deterministic way, giving the impression that the need to use ML is not well justified, which could lead in many cases to an ML misapplication.\par
Moreover, we remarked that in all works that we have covered so far, ML results always appear better than empirical results in the numerical simulation. This fact raises the question whether is it true that ML tools are always outperforming the classical methods or not. At this point, we should mention that we are neither doubting the major success of ML nor questioning its efficiency in solving many problems, instead we are highlighting the fact that in some cases, choosing the data plays an important role in assessing the accuracy of an ML model. To clarify the idea, let's put in place a concrete example. Imagine that you are working on  a computer vision object detection problem and the goal is to detect a UAV in the images. Then, if you do not provide a good quantity of non-UAV images to the model, you will find out that the CNN is providing good accuracy on UAV images and fail in non-UAV images. Moreover, if the test set for example is biased and has some similarity with the training set, you will end up with a good accuracy, but in reality, the model will fail to predict new unseen examples. Hence, the quality and the quantity of the data plays a big role in assessing the accuracy of the model, and neglecting this point will lead to a fake ML accuracy.\par
In the same context, we know that data plays an indispensable role as the learning algorithm is used to discover and learn knowledge or properties from the data. That is why we strongly believe that the  wireless communication community should accredit more importance to providing open source high quality data as we remarked that there is not a sufficient quantity of data online dedicated for wireless purposes compared to the amount of data available for computer vision tasks, for example.\par
Another ML drawback is remarkable in some works where methods are compared in terms of performance. For instance for CNN architecture comparison, you may notice that when comparing two ML models, there is no mathematical explanation as to why such model is better than the other one or why such a NN architecture outperforms another NN architecture. This point illustrates the ``black box`` aspect of ML, in other words, it is a matter of  tuning parameters and evaluating the result and no further explanation could be provided. Consequently, for a given problem we sometimes cannot predict which model will perform well and which model is not promising.\par 
However, ML remains an interesting alternative and a  promising tool for UAV-related problems in particular. Therefore, we believe that several ideas can be addressed in the future. In fact, more complex ML models can be tested on some UAV related problems, for example, in path loss prediction more regression tools can be tested on this problem.
Also for UAV detection problems, we noticed that it is solved either via sound detection or via image detection by converting it into a computer vision problem. Instead, we think that a complex hybrid system that uses different types of inputs (e.g. sound FFT, image, radio transmission.) is feasible by ML, where an adequate NN (e.g. a type of CNN for images and a given type of RNN for sound and radio)  will provide a score for each type of input and then a final NN is used to classify the output.\par
To conclude, ML supervised and unsupervised frameworks have often successfully overcome  many challenges by providing intelligent solutions for various problems involving UAVs.

\section{Reinforcement learning solutions for UAVs}\label{sec:RL}
\subsection{RL Overview}
\begin{figure}
	\centering
	\includegraphics[width=3.5in]{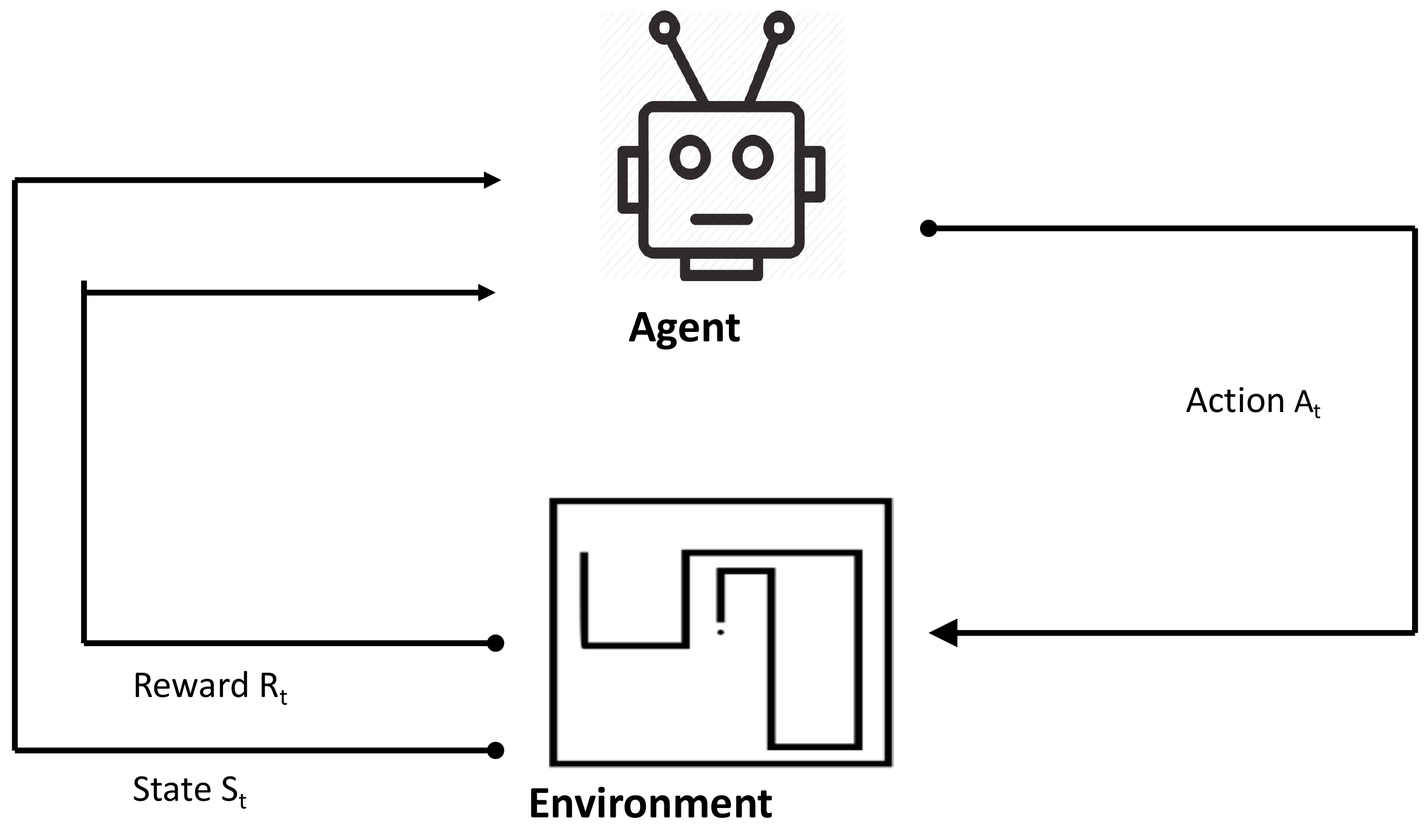}
	\caption{Reinforcement learning elements.}
	\label{fig:reinforcement learning architecture}
\end{figure}
Like the supervised and unsupervised learning areas of ML, RL is another area of ML dedicated to making decisions in a well-defined environment.
Formally, a reinforcement problem always has 5 main elements as shown in the Fig~.\ref{fig:reinforcement learning architecture}:\\
\begin{enumerate}
	\item \textbf{The agent}: An entity that can take an action denoted by $A_t$ and receives a reward $R_t$ accordingly. 
	\item \textbf{The environment}: A representation of the real world in which the agent operates. 
	\item \textbf{The policy}: It is the mapping of each state $S_t$ to an action $A_t$. We usually denote a policy by $\pi$. 
	\item \textbf{The  reward signal}:  The feedback that the agent receives after performing an action. It is denoted on the Fig.~\ref{fig:reinforcement learning architecture} by $R_t$.
	\item \textbf{The value function}: It represents how good a state is, hence it is the total expected future rewards starting from a given state. A value function is usually denoted by $V(s)$ where $s$ is the state that we are interested in. Mathematically, it is formulated as follows: 
	$V(s)=\mathbf{E}(G_t)$, where $G_t$ is the discounted sum of  future rewards: $G_t=\sum_{t} \gamma^{t-1} R_t$ , $\gamma \in [0,1]$.
	
\end{enumerate}

The goal is to decide correct actions (or policy) in a way that maximizes a predefined reward function, which should be adapted to the type of the problem. 
In addition to the 5 elements of RL mentioned above, another element can be considered in some cases, which is the model. Depending on its presence or not, RL problems can be divided into two main categories which are the model-based RL and the model-free RL.  In what follows, we differentiate between these two areas. 
\subsubsection{Model-based RL}
As its name indicates, the model-based RL problem uses a model as a sixth element to mimic the behavior of the environment to the agent. Consequently, the agent becomes able to predict the state and the action for time T+1 given the state and the action at time T. 
At this level, supervised learning could be a powerful tool to do the prediction work. 
Thus, unlike the model-free RL, in model-based RL, the update of the value function is based on the model and not on any experience.
\subsubsection{Model-free RL}
In model-free RL problems, the agent cannot predict the future and this is the main difference with the model-based RL framework explained previously. Instead, the actions are based on the so-called ''trials and errors'' method where the agent, for instance, can search over the policy space, evaluate the different rewards, and pick finally an optimal reward. A well known classic example for model-free RL is the Q-learning method where it estimates the optimal Q-values of each action and reward and chooses the action having the highest Q-value for the current state. 
To summarize, differentiating between model-based and model-free RL problems is an easy task. Just ask yourself the following question: Is the agent able to predict the next state and action? If the answer is yes then you are dealing with a model-based RL, otherwise, it is more likely a model-free RL problem. 

\subsubsection{Deep Reinforcement Learning (DRL) Overview}

While classical RL proposed an efficient solution for many types of discrete decision problems, more realistic solutions could be provided using DRL which has proven its efficiency by reaching super human level control.
DRL is based on using ANN to evaluate action values using the previous experiences of the agent. Among the many algorithms proposed in the literature, in the following section, we go over the most used ones.  For the reader convenience, the two algorithms, Deep Q Network (DQN) and Deep Deterministic Policy gradient (DDPG), are going to be covered briefly. Thus, we kindly refer readers interested in their deep technical details to~\cite{RL35} for DQN and to~\cite{deepmind} for DDPG.  

\textbf{
	Deep Q Network (DQN):
}
DQN was the first algorithm proposed in the context of DRL by Mnih et al. in~\cite{RL35}. To understand the key concepts of DQN, a basic knowledge of Q-learning algorithm is recommended, hence we refer interested readers to Sec.~\ref{sec:Q-learning}. It is worth to mention that DQN was proposed as an improvement to Q-learning which uses a discrete state and action space in order to build the Q-table. In contrast, the Q-values of the DQN are approximated using ANN by stocking all the previous agent experience in a dataset and then feeding it to the ANN to generate the actions based on minimizing a predefined loss function derived from the Bellman equation. We should also mention the fact that the idea of DQN was inspired from Neural Fitted Q-learning (NFQ) proposed in 2005~\cite{RL45} that was suffering from overestimation problems  and instabilities in the convergence. There exist many other improved variations of DQN such as  double DQN, dueling DQN, and distributional DQN. 
Despite the remarkable success of DQN, especially when it was historically tested on ATTARI games, it has its own limitations such as the fact that it cannot deal with continuous space action and cannot use stochastic policies.\par
\textbf{Deep Deterministic Policy Gradient (DDPG): }
\label{DDPG}
To overcome the restriction of discrete actions, Deterministic Policy Gradient (DPG) algorithm was first proposed   in Deepmind's publication in 2014~\cite{deepmind} based on an Actor-Critic off policy approach. We refer readers that are not familiar with Actor-Critic RL methods to~\cite[Chapter 13]{RLbook}. For the sake of simplicity, let's keep in mind that  Actor-Critic methods are generally composed mainly of two parts: a Critic that estimates either the action-value or the state-value and an Actor that updates the policy in the direction proposed by the Critic. Later on, in 2015, and based on the DPG algorithm, Deepmind proposed a new DRL algorithm called the Deep Deterministic Policy Gradient (DDPG) algorithm.  DDPG is a model-free, off-policy method that is based on Actor-Critic algorithm. 
In short, DDPG is a DRL algorithm that helps the agent to find an optimal strategy by maximizing the reward return signal. The main advantage of such deep algorithm is that it performs well on high-dimensional/infinite  continuous action space. 

\subsection{Case Study}\label{SEC:CaseStudy}

\begin{figure}
	\centering
	\centerline{\includegraphics[width=3.4in]{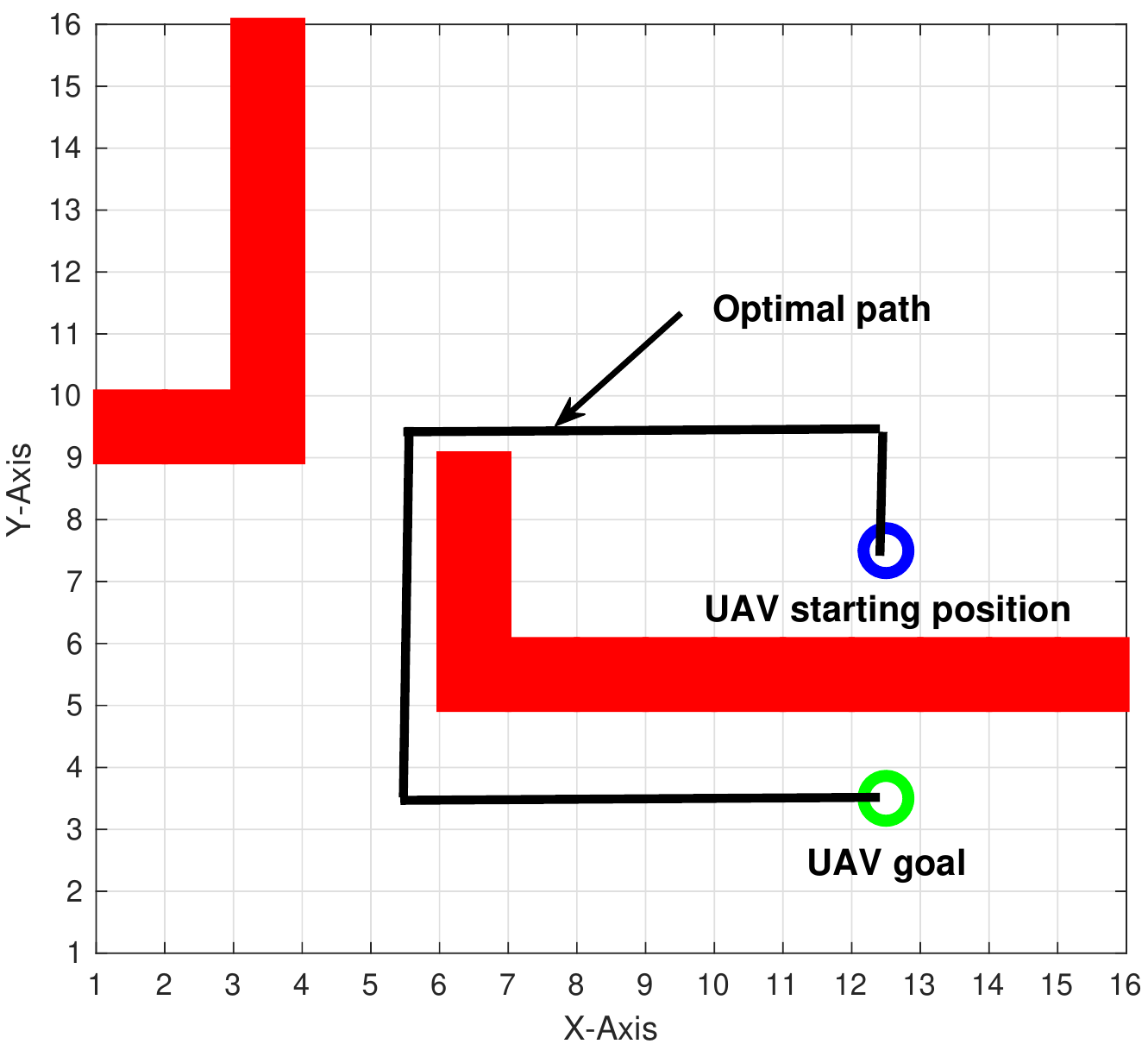}}
	\caption{Grid map.}
	\label{fig:grid}
\end{figure}
Motivated by its popularity among RL algorithms, we introduce Q-learning which is a classical free-model RL algorithm. We intend to provide a comprehensive and practical explanation to the reader on how RL could be used in UAV path planning problem. Readers with a basic knowledge on RL could definitely skip this section. We stick to a basic example where a UAV flying at a fixed altitude learn how to reach a given target while avoiding obstacles in the map shown in Fig.~\ref{fig:grid}.\par

\subsubsection{Q-learning overview}
\label{sec:Q-learning}
Q-learning algorithm is based on the Q-table used to select actions for the agent at each step. The table is composed of the combination of every state with every possible action and hence its dimension is $|States|\times |Actions|$. The Q-table is used to store and update the maximum future reward referred to by Q($state_i,action_j$) which is the $(i_{th},j_{th}) $ entry of the Q-table. This Q-table is of great importance to the Q-learning algorithm simply because it is used to determine which action should the agent take such that the expected future reward is maximized. 
\subsubsection{Update rule}
The update of the Q-table is done using a fundamental equation in RL which is the Bellman equation: 

\begin{gather} 
\label{eq:lab}
\scalebox{.8}{$Q_{new}(s_t,a_t)=(1-\alpha)Q_{old}(s_t,a_t)+\alpha(R_{t+1}+\gamma max_a(Q(s_{t+1},a)),$}
\end{gather}
where $s_t,a_t$ are respectively the state and the action taken at time t, $\alpha$ is the learning rate, which allows the old value of the Q-table to influence current updates, $\gamma$ is the discount factor, which is a measure of how future rewards will affect the system. After every taken action, the agent updates its Q-table values using~(\ref{eq:lab}), then, at a given state, it selects the action having the highest Q-value.
\subsubsection{The exploration/exploitation dilemma}
\begin{figure}
	\centering
	\includegraphics[width=3.6in]{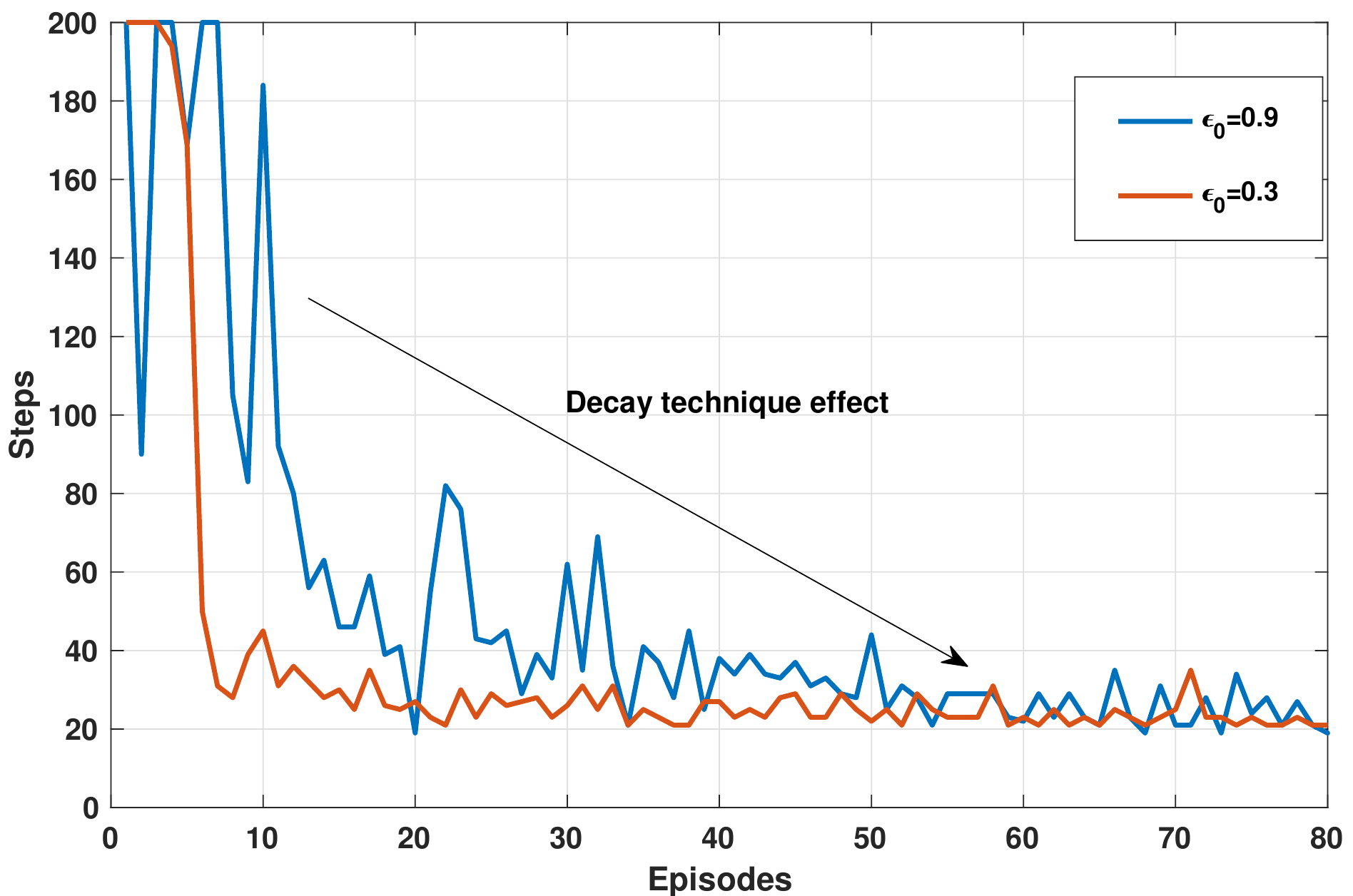}
	\caption{Exploration/Exploitation dilemma.}
	\label{fig:EXP}
\end{figure}

One fundamental concept for RL, which is visible also in Q-learning, is the exploration/exploitation dilemma. To explain this duality, let's discover how the agent will succeed in reaching its goal. First, the agent makes a random step in the environment, then it starts updating the Q table (initialized with zeros for example) according to~(\ref{eq:lab}). However, if the agent only uses the Bellman equation, it is possible that it is stuck in a good state forever, while better states exist on the map. It is similar to a case of an optimization process that is stuck in a local minimum or maximum while better solutions still exist by exploring the environment. To solve the last problem, the exploitation/exploration dilemma is proposed. This duality introduces a randomness into the system so that the agent at each step could either exploit the environment by selecting actions that maximize the Q-values of the Q-table, or explore the system by selecting some  random actions.  The parameter that usually refers to the probability threshold for exploration is denoted by $\epsilon$. In our implementation, we used a decay technique that decreases the value for epsilon at each episode so that we encourage exploration at the beginning of the process, usually known as early exploration, and then prioritize exploitation so that the agent can use the learned paths. Fig.~\ref{fig:EXP} shows the effect of the initial value of $\epsilon$, denoted by $\epsilon_0$, on the convergence of the system. The red line, corresponding to a low $\epsilon$ value, converges more rapidly since the exploration probability is low and hence the system will rapidly use the optimal values from the Q-table to take actions. However, it is clear that for the blue line using early exploration, additional  randomness is introduced at the beginning of the process due to the high starting value of $\epsilon$. We also remark that the number of steps is decreasing due to the fact that the UAV has already found its optimal path, shown by a solid black line in Fig.~\ref{fig:grid}, starting from episode 20 approximately. 		

RL is considered a promising framework for the UAV network in many scenarios, in the following we cover part of its applications based on the existing literature

\subsection{RL Solutions for UAV-based Problems}

\subsubsection{RL for autonomous navigation
}

A network of UAVs can no longer be controlled in a classical way by manually controlling the navigation of each UAV from the network separately. It is highly recommended nowadays to equip UAVs with the ability to make intelligent decisions by implementing a high level of control. Achieving such a high autonomy for UAVs is a challenging task due to the continuous changes in the UAVs environment and to the different constraints related to UAV navigation (e.g. battery, UAV dynamics).
\par In~\cite{RLtable1}  a high-level control method of UAVs is implemented for uncertain or unknown environments. Although model-based algorithms are generally not suitable for real-time applications due to the expensive computation needed for learning the model and deciding actions to take, the paper uses a modified model-based RL by implementing Texplore algorithm te perform path planning task for a UAV. The advantage of Texplore is that it separates the action selection, model learning, and planning by performing them in a parallel manner.
Simulation results show that Texplore algorithm outperforms the classical Q-learning method by avoiding exhaustive exploration of the environment.
The work done so far is interesting but still limited to a 2D problem with a simplified map and hence could be extended in the future to a more complex and challenging 3D environment where the UAV can adjust its height in order to avoid potential obstacles.\par

Beyond pure path planning, the work in~\cite{RLtable3} investigates providing coverage for ground users by studying the deployment and the trajectory design of a network of UAVs in order to meet several performance metrics such as; coverage, minimum interference, and best quality of experience (QoE). The proposed model-free RL-assisted framework enables dynamic tracking of users' movement by adjusting the UAV's location accordingly. 
The idea starts by clustering the users using a classical GAK-means algorithm ( which is a modified version of the  $K$-means algorithm explained briefly in Sec.~\ref{sec:SuperAndUnsuper}). It is worth mentioning that deploying the UAV at the cluster center does not meet optimality simply because the performance metrics adopted in the problem are not only related to the euclidean distance to users, instead they are related to other parameters such as the altitude of the UAV and the  LOS presence. Consequently, a Q-learning algorithm is proposed to firstly deploy UAVs in a sub-optimal way and decide their trajectory later.
The work done so far is of great importance, however, some assumptions made may be far from reality, such as the fact that users, when moving around, are not supposed to mix with other clusters. Moreover, positioning the UAVs initially using the $K$-means results might be better than selecting a random location in terms of fast convergence to the sub-optimal positions.\par
More other works in the literature have  focused on providing coverage for ground users. Since the problem setup is composed of several UAVs, some of these works investigated using multi-agent RL techniques.  For instance,  in~\cite{RL34} optimal coverage is studied through a distributed RL algorithm based on Multi Agent Reinforcement Learning (MARL) which consist of applying RL techniques for the type of problems where multiple agents are operating in the same environment. Each agent has the objective of maximizing its own future rewards by interacting with the environment. Moreover, authors in~\cite{RLtable9} studied using multi-agent Q-learning to design a trajectory for a network of UAVs. The objective is to maximize the information rate while minimizing the power consumption. In addition to the multi-agent Q-learning algorithm that was used to determine the optimal positions for UAVs, an echo state network was used to predict the users' movement. In~\cite{RLtable11}, the author proved that double Q-learning could be an improvement for the classical Q-learning for problems related to wireless coverage. A new framework for trajectory design was proposed in order to maximize the number of users having a good QoE. 
\par So far, we covered classical RL solutions for UAVs path planning problems, however, 
a more complex autonomous navigation solutions could be provided via DRL. In what follows, we go over the most relevant works coupling DRL with autonomous navigation for UAVs. 

In the context of providing coverage for ground users, DRL can play an important role in building efficient solutions.  In such a setup, the UAVs are usually deployed as flying base stations or relays. For instance, the authors in~\cite{RL12} investigated applying DRL to a multiple input multiple output (MIMO)-based UAV network where each UAV is equipped with a single antenna.  The proposed DRL solution is based on DQN where the Signal-to-Interference-Plus-Noise Ratio (SINR) was used as a metric for the quality of the channel and based on which the reward signal is defined.  The UAV maximizes the expected reward calculated based on the received signal strength. Consequently, the UAV maximizes its coverage efficiency measured based on a predefined coverage score. The proposed solution was finally compared with other DRL methods and proved its superiority in some setups. However, we think that comparing it to a different type of DRL algorithms such as DRL-based energy efficient control for coverage and connectivity (DRL-EC$^{3}$) proposed in~\cite{RL13} is not a fair comparison since the latter solution is based on DDPG where the action space and the state space are continuous. In addition, the state space considered in the  solution provided in~\cite{RL12} is limited to 3 possible cases related to the received signal strength.\par
In addition to providing ground wireless connectivity, there exist a plethora of areas where UAVs could be used efficiently, such as drone delivery. In this frame, achieving drone delivery tasks through DRL was investigated in~\cite{RL11}. The authors used double DQN to propose a path planning algorithm for UAVs having an objective to reach a destination in an obstacle-impaired environment. The proposed solution is an improvement to the authors' previous work in~\cite{RL36}, where three DRL algorithms were tested which are the DQN, double DQN, and duel DQN. As double DQN gave the best results, in~\cite{RL11} the same algorithm was used and the depth information deduced from the image of the UAV stereo-vision front camera was used as an input. 
More futuristic ideas are discussed in the literature, such as using drones to serve food and drinks in restaurants~\cite{RL7}.\par 

To this point, we have mainly focused on research works that are based on Q-learning methods, either using a classical RL or DQN. However, policy gradient methods could be applied to a wider range of RL problems. For instance, DDPG algorithm, which belongs to the set of policy gradient methods, is more suitable for complex problems, especially when dealing with continuous action space. To make it simpler, let's consider a UAV path planning problem. To apply DQN  you need to discretize the action space and accordingly, the UAV will have a well-defined set of movement directions that it can perform. The discretization process can be sometimes computationally expensive and, in other cases, it is even impossible to implement it, especially when targeting real-time applications. As a solution, policy gradient methods could be easily implemented
to perform continuous action. For instance, in the context of UAV motions, actions could be related to the speed values and direction angles. It is important to point out that policy gradient methods are not always better than Q-learning-based methods and that they have their own drawbacks such as the high variance problem in estimating the expectation of the reward.
In what follows, we are going to cover some of the relevant works that apply this type of methods to solve path planning tasks for UAVs.\par	
Back to providing wireless connectivity for ground users, the authors in~\cite{RL13} used an Actor-Critic based method to solve a multi-objective control problem where the UAVs tend to minimize their energy consumption and maximize their coverage range. Unlike the previously discussed DQN-based solutions, this work is based on continuous action space formed by the UAV direction and the flying distance for each UAV. Moreover, the authors take into consideration coverage fairness which is an important indicator since maximizing coverage could fall into covering a small subset of ground users. As a solution for the defined multi-objective problem, the authors adjusted the DDPG algorithm accordingly and called it DRL-EC$^{3}$. The new algorithm was compared to two baseline methods and proved its superiority in terms of coverage score and energy consumption.

In~\cite{RL4} the environment  considered is  a complex large scale three-dimensional map. In other words, the map is crowded with obstacles, where all directions are possible for the UAV, and it is also dynamic.  Those types of maps are quite challenging for RL path planning due to the fact that it is very difficult to rely on methods that use maps to represent the environment for the agent. The solution proposed is based on modeling the navigation problem using  partially observable Markov decision process (POMDP) and then solving it by applying a DRL-based algorithm called Fast-RDPG. It is worth to mention that 1) POMDP is an extension of Markov decision process (MDP) and that 2) recurrent deterministic policy gradient algorithm (RDPG) belongs to another set of DRL algorithms that are based on DPG.

In~\cite{RL5}, DDPG algorithm, briefly introduced in Sec.~\ref{DDPG}, is used to train the UAV to navigate in a 3D environment while avoiding obstacles. The proposed solution considers a continuous action space which explains the use of a DDPG-based approach. The authors used transfer learning to accelerate the learning by using the weights learned by the UAV after being trained in a free space environment. An urban environment is simulated by adding obstacles in specific locations and penalizing the UAV for any crash that occurred while reaching its target location. Numerical results showed that the UAV succeeded in reaching its target while avoiding all obstacles in the way. However, the success rate of the UAV decreases with the complexity of the environment by adding more obstacles. The lacking of precision was explained by the fact that using infinite continuous  action space make it hard to reach full accuracy.\par   
DDPG was also used differently in~\cite{RL10} to jointly design a path for a network of UAVs in order to maximize its throughput. The idea proposed is to formulate the problem as an MDP where the reward is related to the throughput and the constraints are related to total transmission power and channel availability. The actions taken by the UAV are related to adjusting both the 3D location and the transmission control. Due to the fact that the actions and the states are continuous, DDPG framework was used in 3 three different setups. For each setup, the reward function is changed to achieve a given control objective.\par 
Several other works existing in the literature are quite interesting. For instance, environment exploration and  obstacle avoidance problems for UAVs are solved via different RL methods with both continuous and discrete space action in~\cite{RL16,RL17,RL18,RL19,RL20,RL21,RL22,RL23,RL24,RL25,RL26,RL27}. In~\cite{RL37} the optimal deployment of UAVs that minimizes several parameters such as transmission power, caching, and the number of UAVs, is achieved through RL.  Other works tested RL for assisting a UAV in a landing operation~\cite{RL29,RL30,RL31,RL32}. In~\cite{RL33} an anomaly detection is performed via RL in order to detect abnormality in the functioning of the motor and launch the landing procedure immediately.

All the previously discussed works  only focus on a specific type of RL application which is path planning. The  Table.~\ref{Tab:table2}  summarizes all works in literature up to the date of writing this survey. The cited works are oriented towards trajectory planning solutions for UAV-based networks. Most of the cited works are recent and are classified according to the application main objective and to the different parameters used. We insist on the fact that the classification of dimensionality in 3D and 2D works is made on the basis of the movement of the UAV and not on the nature of the environment, in other words, a work where a 3D environment is considered and where the UAV flies at a fixed altitude is classified as a 2D work. In addition, the classification according to battery level parameters is marked if the UAV's limited energy is considered as a constraint to the problem, so any work where the UAV's energy consumption is minimized without imposing a limit on the battery level will not be marked.  In the coming section, we will cover more interesting potential applications of RL such as event scheduling and resource allocation.\par

\begin{table*}[]
	\large
	\caption{  Reinforcement learning solutions for path planning UAV-based problems.}
	
\begin{adjustbox}{width=\textwidth,center}
		\label{Tab:table2}
		\def\arraystretch{1.5}
		\begin{tabular}{|c|P{5cm}|P{5cm}|c|c|P{5cm}|c|P{2cm}|c|P{2cm}|P{2cm}|}
			\hline
			
			&                                &                               &                            &                                                                                                           & \multicolumn{6}{c|}{Problem Parameters}                                                                                                            \\ \cline{6-11} 
			\multirow{-2}{*}{Paper} & \multirow{-2}{*}{RL technique}  & \multirow{-2}{*}{Application Objective} & \multirow{-2}{*}{3D/2D} & \multirow{-2}{*}{\begin{tabular}[c]{@{}c@{}}Single/Multiple\\ UAVs\end{tabular}} & \begin{tabular}[c]{@{}c@{}}Wireless communication\\ based parameters\end{tabular} & Obstacles & LOS/NLOS & Battery & UAV dynamics & Users movement \\ \hline \multicolumn{11}{|c|}{\textbf{Discrete action space (Classical RL)}}  \\ \hline  
			\cite{RLtable1}&  Q-Learning based Texplore                                & Reaching a target                        &2D                        &Single                                                                                 &\xmark                                                                                   & \xmark          & \xmark         &\cmark         & \cmark            &\xmark                 \\ \hline
			\cite{RLtable2}& DRL ESN-Based                                   &  Multiple objective optimization                  &   2D                     & Multiple                                                                                & Interference/ Wireless latency/ Transmit power
			& \xmark           &   \xmark        &   \xmark       & \xmark              &            \xmark    \\ \hline
			\cite{RLtable3}	&	K-means+Q-learning         & Ground coverage                    &  3D                            &Multiple                        &QoE at the users                                                                                 &       \xmark                                                                             &    \xmark                   &\xmark          &       \xmark        &     \cmark            \\ \hline
			\cite{RLtable9}	&Q-learning+ESN+GAK-means/ Power control   & Ground coverage  &  3D                            &Multiple                        &Information rate and transmit power                                                                                 &       \xmark                                                                             &    \cmark                   &\xmark          &       \xmark        &     \cmark            \\ \hline
			
			\cite{RLtable5}&  Q-learning vs NN-based Q-learning
			& Ground coverage                            &2D                        &  Single                                                                               &   Transmission rate 
			&Single           & \cmark         & \xmark        &   \xmark            & \xmark                \\\hline
			
			\cite{RLtable10}&  Q-learning
			&   Ground coverage  &2D                        &  Single                                                                               &   QoE at the users 
			&Single           & \xmark         & \cmark        &   \xmark            & \xmark                \\ \hline
			\cite{RLtable6}&     Q-learning		                            &  Obstacle avoidance                 &         2D               & Single                                                                                &    \xmark                                                                                &       Multiple    &      \xmark     &  \xmark        &         \xmark      &   \xmark              \\ \hline
			\cite{RLtable11}&     Double Q-learning
			& Ground coverage                          &2D                        &       Multiple                                                                          &  User satisfaction                                                                                 &\xmark          &  \cmark        &     \xmark    &      \xmark        &   \xmark             \\ \hline
			\cite{RLtable12}&     Q-learning
			& Ground coverage                         &3D                        &       Multiple                                                                          &  Throughput/Interference                                                                                 &Multiple          &  \cmark        &     \xmark    &      \xmark        &   \cmark             \\ \hline
			\cite{RLtable14}&     S-MGD
			&  Ground coverage                     &3D                        &       Multiple                                                                         &  Downlink                                                                                &\xmark          &  \cmark        &     \cmark    &      \xmark        &   \cmark             \\ \hline
			\cite{RLtable15}&     Enhanced Multi agent Q-learning
			& Sensing task                     &3D                        &       Multiple                                                                         &  \xmark                                                                                &\xmark          &  \cmark        &     \xmark    &      \xmark        &   \xmark             \\ \hline
			\cite{RLtable16}&     Multi agent RL (MAXQ) 
			& Collision avoidance                    &3D                        &       Multiple                                                                         &    \xmark &Multiple          &  \xmark        &     \xmark    &      \cmark        &   \xmark             \\ \hline
			\cite{RLtable19}&     Q-learning
			&  Maximize ground user throughput                  &2D                        &       Single                                                                         &    NOMA throughput &Multiple          &  \cmark        &     \xmark    &      \xmark        &   \cmark             \\ \hline
			
			\cite{RLtable22}&     Q-learning
			&    Collision avoidance               &2D                        &       Multiple                                                                         &    Path loss/ SNR &Multiple          &  \xmark        &     \xmark    &      \xmark        &   \xmark             \\ \hline
			\cite{RLtable27}&    enhanced multi-agent Q-learning
			&  Target localization       &2D                       &      Multiple                                                                        &    Channel modeling &\xmark          &  \cmark        &     \cmark    &      \cmark        &   \xmark             \\ \hline		
			\cite{RLtable34}&    Sarsa vs Q-learning 
			&  Data collection       &2D                       &      Single                                                                      &   Channel gain/ Data rate &\xmark          &  \xmark        &     \xmark    &      \xmark        &   \xmark             \\ \hline	   
			\multicolumn{11}{|c|}{\textbf{Discrete action space (DRL)}}  \\ \hline 
			\cite{RLtable7}&    DQN
			&    Reaching a target                       &3D                        &   Single                                                                              &    \xmark                                                                               &  Multiple         & \xmark         &   \xmark      &   \xmark           &    \xmark            \\ \hline
			\cite{RLtable13}&     DQN
			&       Ground coverage                    &3D                        &       Single                                                                         &  Spectral efficiency                                                                                 &\xmark          &  \cmark        &     \xmark    &      \xmark        &   \xmark             \\ \hline
			\cite{RL12}&     DQN
			& Ground coverage &2D                        &       Multiple                                                                        &  MIMO technique/ interference                                                                                 &\xmark          &  \xmark        &     \xmark    &      \xmark        &   \xmark             \\ \hline
			\cite{RLtable21}&     Multi agent DQN
			& QoS at users                        &3D                        &       Multiple                                                                         &    SNR/Throughput &\xmark         &  \cmark        &     \xmark    &      \xmark        &   \xmark             \\ \hline
			\cite{RLtable23}&     Q-learning+DQN
			&  Ground coverage                  &2D                        &       Both                                                                         &    Path loss/ Interference/ Coverage and fairness score  &\xmark          &  \cmark        &     \cmark    &      \xmark        &   \xmark             \\ \hline
			\cite{RLtable29}&     DQN
			&  Data collection                  &2D                        &       Single                                                                       &    Path loss/ SNR  &\xmark          &  \xmark        &     \xmark    &      \xmark        &   \xmark             \\ \hline
			\cite{RLtable30}&     DQN
			&  Combat mission                  &3D                        &       Single                                                                       &   \xmark  &\xmark          &  \xmark        &     \xmark    &      \cmark        &   \xmark             \\ \hline
			\cite{RLtable31}&    Decaying DQN
			&  multi-objective optimization              &2D                        &       Single                                                                       &   NOMA channel modeling/Bandwidth/SINR  &\xmark          &  \cmark        &     \xmark    &      \xmark        &   \xmark             \\ \hline
			\cite{RLtable32}&    Mutual DQN
			&  Cellular offloading           &3D                       &       Multiple                                                                       &   NOMA signal modeling/Data rate/Interference  &\xmark          &  \cmark        &     \xmark    &      \xmark        &   \cmark             \\ \hline
			\cite{RLtable36}&   DQN
			&  Reaching Target      &3D                       &      Single                                                                     &   \xmark &\xmark          &  \xmark        &     \xmark    &      \cmark        &   \xmark             \\ \hline	
			\cite{RLtable37}&   DQN
			& Pursuer-evader problem     &3D                       &      Single                                                                     &   \xmark &Multiple          &  \xmark        &     \xmark    &      \cmark        &   \xmark             \\ \hline
			\cite{RLtable39}&   Double DQN
			& Cellular offloading     &2D                       &      Single                                                                     &   \xmark &\xmark          &  \xmark        &     \cmark    &      \xmark        &   \cmark             \\ \hline
			\multicolumn{11}{|c|}{\textbf{Continuous action space (DRL)}}  \\ \hline                      
			\cite{RLtable8}&     DDPG
			&  Ground target tracking                           &2D                        &       Single                                                                          &  \xmark                                                                                 &Multiple           &  \cmark        &     \xmark    &      \cmark        &   \xmark             \\ \hline
			\cite{RL5}&     DDPG
			&  Reaching a target                          &3D                        &       Single                                                                          &  \xmark                                                                                 &Multiple           &  \xmark        &     \xmark    &      \cmark        &   \xmark             \\ \hline
			\cite{RL13}&     DDPG (DRL-EC)
			&   Multi-objective optimization                       &2D                        &       Multiple                                                                          &  Coverage score/Energy                                                                                 &\xmark           &  \xmark        &     \xmark    &      \xmark        &   \xmark             \\ \hline
			\cite{RL4}&    Fast-RDPG
			&Obstacle avoidance                         &3D                        &       Single                                                                          &           \xmark                                                                      &Multiple           &  \xmark        &     \xmark    &      \cmark        &   \xmark             \\ \hline
			\cite{RLtable17}&     DRL (LwH)
			& Reaching a goal &2D                        &       Single                                                                         &    \xmark &Multiple          &  \xmark        &     \xmark    &      \cmark        &   \xmark             \\ \hline
			\cite{RLtable18}&     Compound action Actor-Critic (CA2C)
			&  Cooperative sensing         &2D                        &    Multiple                                                      &    Transmit power/ SNR/ Data rate &\xmark          &  \cmark        &     \xmark    &      \xmark        &   \xmark             \\ \hline
			\cite{RLtable20}&     Multi agent DDPG
			&  Fair ground coverage       &2D                        &      Multiple                                                                         &    Data rate &\xmark          &  \xmark        &     \xmark    &      \xmark        &   \xmark             \\ \hline
			\cite{RLtable24}&    DRL (Actor-Critic)
			&  Long-term coverage       &2D                        &      Multiple                                                                         &    Coverage fairness score &\xmark          &  \xmark        &     \xmark    &      \xmark        &   \xmark             \\ \hline		
			\cite{RLtable25}&    DDPG
			&  Network utility       &3D                       &      Single                                                                         &    SINR / Path loss &\xmark          &  \cmark        &     \xmark    &      \cmark        &   \xmark             \\ \hline		    		   
			\cite{RLtable26}&    DDPG
			&  Obstacle avoidance       &3D                       &      Single                                                                         &    \xmark &Multiple          &  \xmark        &     \xmark    &      \cmark        &   \xmark             \\ \hline		    		   
			
			\cite{RLtable28}&    DDPG (EEFC-TDBA)
			&  Ground coverage       &3D                       &      Single                                                                        &    Bandwidth/Data rate/Throughput &\xmark          &  \cmark        &     \cmark    &      \cmark        &   \xmark             \\ \hline		    	                          
			\cite{RLtable33}&    DDPG
			&  Ground coverage       &2D                       &      Multiple                                                                      &   Data rate/ Bandwidth &\xmark          &  \xmark        &     \cmark    &      \xmark        &   \xmark             \\ \hline		
			\cite{RLtable35}&   jPPO+ConvNTM
			&  Data collection       &2D                       &      Multiple                                                                      &   \xmark &\cmark          &  \xmark        &     \cmark    &      \xmark        &   \xmark             \\ \hline		                    
			\cite{RLtable38}&   Multi agent DDPG
			& Ground coverage    &3D                       &      Multiple                                                                    &   Path loss/Bandwidth/SINR & \xmark          &  \cmark        &     \xmark    &      \xmark        &   \xmark             \\ \hline				                    
			
		\end{tabular}
\end{adjustbox}
\end{table*}
\subsubsection{RL for IRS-enabled UAV Networks}
\begin{figure}
	\centering
	\includegraphics[width=3.5in]{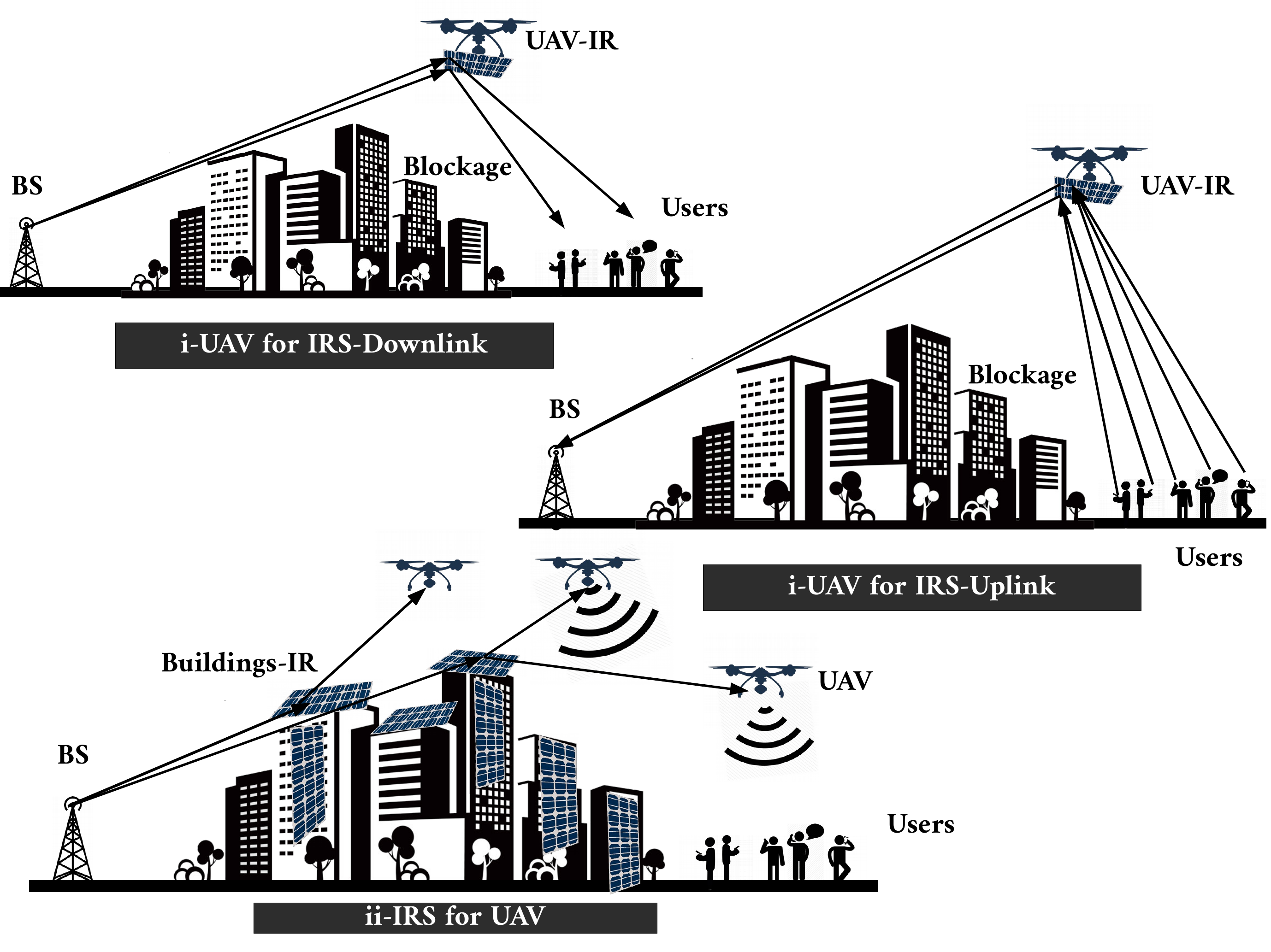}
	\caption{UAV-IR use case scenarios.}
	\label{fig:IRS}
\end{figure}

In this section, we cover a promising technology for 6G networks through the use of intelligent reflective surfaces (IRS). This next-generation technology consists of using a set of passive reflective elements to control the wireless propagation and direct the signal to a target direction. In recent literature, some works have considered equipping UAVs with IRSs to increase spectrum efficiency. Other works have investigated the use of the IRS to reflect the signal towards UAVs that are acting as a flying base stations. In Fig.~\ref{fig:IRS}, we visualize these two different scenarios where the IRSs are coupled with UAVs: (i) UAVs for IRSs, where the UAVs carrying the reflective surfaces can act as a passive relay in both uplink and downlink communications between terrestrial BSs and ground users and (ii) IRSs for UAVs, where the IRS-equipped buildings assist the UAV's communication.  
In what follows, we review the work done for these different scenarios, including support for millimeter wave (mmWave) communications.\par
In the context of supporting mmWave technology, the authors in~\cite{RL46} propose an efficient deployment for the IRS-equipped UAV to serve a moving ground user that does not have a LOS with the terrestrial base station. In addition to reflecting the mmWave signals, IRS is also used to harvest energy at the UAV and Deep Q network is used to set the UAV location and the reflecting parameters of the IRS so that the dowlink capacity is maximized. The same authors in~\cite{RL47} consider overcoming the blockage induced by the buildings through IRS-equipped UAV but for multiple users. To this end, distributional RL is used to optimize the UAV's location, the precoding matrix at the base station and the reflection coefficients.\par
Unlike~\cite{RL46}, the work in~\cite{RL48} does not take into account energy harvesting, but rather focuses on minimizing the UAV energy consumption. The authors study the improvement of the quality of service of a UAV-carried intelligent reflector (UAV-IR) by optimizing its location, the phase shift of the IRS and its power allocation to mobile users on the ground. To further improve the efficiency of downlink communications, a non-orthogonal multiple access technique is used. Decaying DQN is used to dynamically adjust the position of the UAV and the algorithm is found to converge and avoid oscillations compared to the classical Q-learning algorithm. Although D-DQN has proven its efficiency compared to the classical Q-learning, it still interesting to see some comparison with other DQN-based algorithms.\par
From an uplink perspective, the authors of~\cite{RL50} study the use of the UAV-IR as a passive relay for the transmission of IoT devices to a terrestrial base station. DRL is used to optimize the UAV location, the IRS phase shift, and the transmission scheduling in order to minimize the average age of the information.\par
Unlike the above works, where IRS-equipped UAVs are studied, in~\cite{RL49}, improving the channel condition between a UAV and a set of ground users is considered by assisting the UAV with IRSs that are mounted on top of the surrounding buildings. In this configuration, several parameters are optimized, such as data throughput, UAV trajectory, and IRS phase shifts. In terms of the reinforcement learning algorithm used, the authors adopt two different types of solutions, a discrete action space solution based on DQN and a continuous action space solution using the DDPG algorithm.\par
\subsubsection{RL for scheduling and resource management
}
Beyond path planning for smart UAVs, one can think about smart event scheduling for a drone network. In this context, the authors in~\cite{RL8} propose a spatiotemporal scheduling framework for autonomous UAVs. 
The proposed RL solution is model-free based and uses the popular Q-learning algorithm.
The algorithm is handling the unexpected events iteratively by checking at every time slot their existence and updating the UAVs schedule accordingly. After that, the trajectory of the UAV is updated according to the Q-learning strategy. There are multiple parameters taken into account for every event (e.g. starting time, processing time, location, priority).
The considered work is interesting for many reasons; it takes into account multiple factors such as dealing with unexpected events efficiently, it also considers the battery level, and works within a cooperative UAV environment. However, it is still not clear how to select some parameters optimally. For instance, the time discretization parameter enables a trade-off between the complexity and time efficiency. In other words, deciding in an optimal way the next event will inevitably result in an increased time processing. This will certainly affect the coverage rate of the UAV badly. Moreover, the author could have considered a more realistic case where multiple docking stations are available instead of considering only one station for the whole network, and therefore the UAV should always consider moving to the nearest station if needed.\par

In~\cite{RL14} a UAV network is managed by ensuring the UAV connectivity given the available bandwidth and energy. The set of drones are charged by a wind-powered station that enables a green wireless power transfer. The number of UAVs authorized to take off is managed through classical RL by solving a system of Bellman optimality equations in order to extract the optimal policy. The authors focused on the physical implementation of the charging station and the drone receiving pads by going through the different technical details of the wireless power transfer system.
Among the assumptions made throughout the work is the fact that the charging time is constant, which could be hampered by several factors.  First of all, while establishing wireless power transfer, the UAV could face a number of  problems such as losing LOS connection with the station or  some misalignment issues. Secondly, the fact that the charging station uses wind power makes it subject to variability in the harvested power. To solve the latter problem, the authors propose using adaptive current control.\par
The authors in~\cite{RL41} investigate the dynamic management of information exchange in a UAV network by studying two different information sharing schemes: the dynamic channel model and the dynamic time slot model. In the first scheme, the channel is shared for the exchange of information in the same time slot, while in the second model, time slots are shared according to the priority level. To solve the problem of dynamic management in the two proposed schemes, the authors used a DQN network coupled with a Long Short Term Memory network (LSTM). The purpose of using the LSTM network is to speed up the convergence of the DQN model since this network helps to predict future states of the environment.\par
In~\cite{RL42}, UAVs are considered as remote edge computing systems that could offload ground users when they are excessively overwhelmed with computation. Hence, the authors studied the best way to assign a UAV to execute a given task in order to minimize the overall mission time. They formulated this resource allocation problem as a Markov decision process and proposed an Actor-Critic based RL technique (A2C) as a solution. In addition to offloading terrestrial users through UAVs, it is possible to perform the inverse and thus offloading the UAVs through Mobile Edge Computing(MEC) server. In this context, the work in~\cite{RL43} investigates offloading the UAV network via MEC servers to minimize the processing time of the drones and save their energy. The complex task of associating each UAV with its corresponding task based on its available energy, and of choosing the optimal MEC server to offload, are carried out on the basis of two Q-learning models. In the same direction, authors in~\cite{RL44} investigate offloading a network of UAVs via MEC where a central control system has the ability to turn the UAV computing units on and off and to decide whether a given task should be performed by another UAV. This management problem is solved through RL techniques.  

Resource allocation is another potential problem that can be solved with RL. The work in~\cite{RL9} is among the  few works that go beyond  UAV deployment or trajectory design, instead, it focuses on resource allocation for a network of multiple UAVs that communicates with ground users. The solution provided is based on Multi Agent Reinforcement Learning (MARL) and the problem formulation is based on stochastic game theory. The authors investigate sub-channel selection, user selection, and also power allocation for each user. Several parameters are taken into account such as the SINR, LOS and NLOS conditions with the users. The work described is of great importance, especially when considering the scarcity of publications in this  particular application. Some other works investigated resource management for UAVs such as in~\cite{RL38} where the number of handoffs that occur when a ground user does not receive its minimal signal from its serving UAV is minimized. The solution used is based on RL. More specifically, it uses the Upper Confidence Bound (UCB) algorithm. In~\cite{RL39} UAV cluster task scheduling problem is addressed by solving the channel allocation problem using DRL. Moreover, the authors in paper~\cite{RL40} propose a comparison between Swarm intelligence and the Q-learning algorithm of RL. The objective is to enable autonomous swarm coordination for a network of high altitude platform stations.
\subsection{
	Discussions and Future Works: 
}

\begin{figure*}
	\centering
	\includegraphics[width=0.8\textwidth]{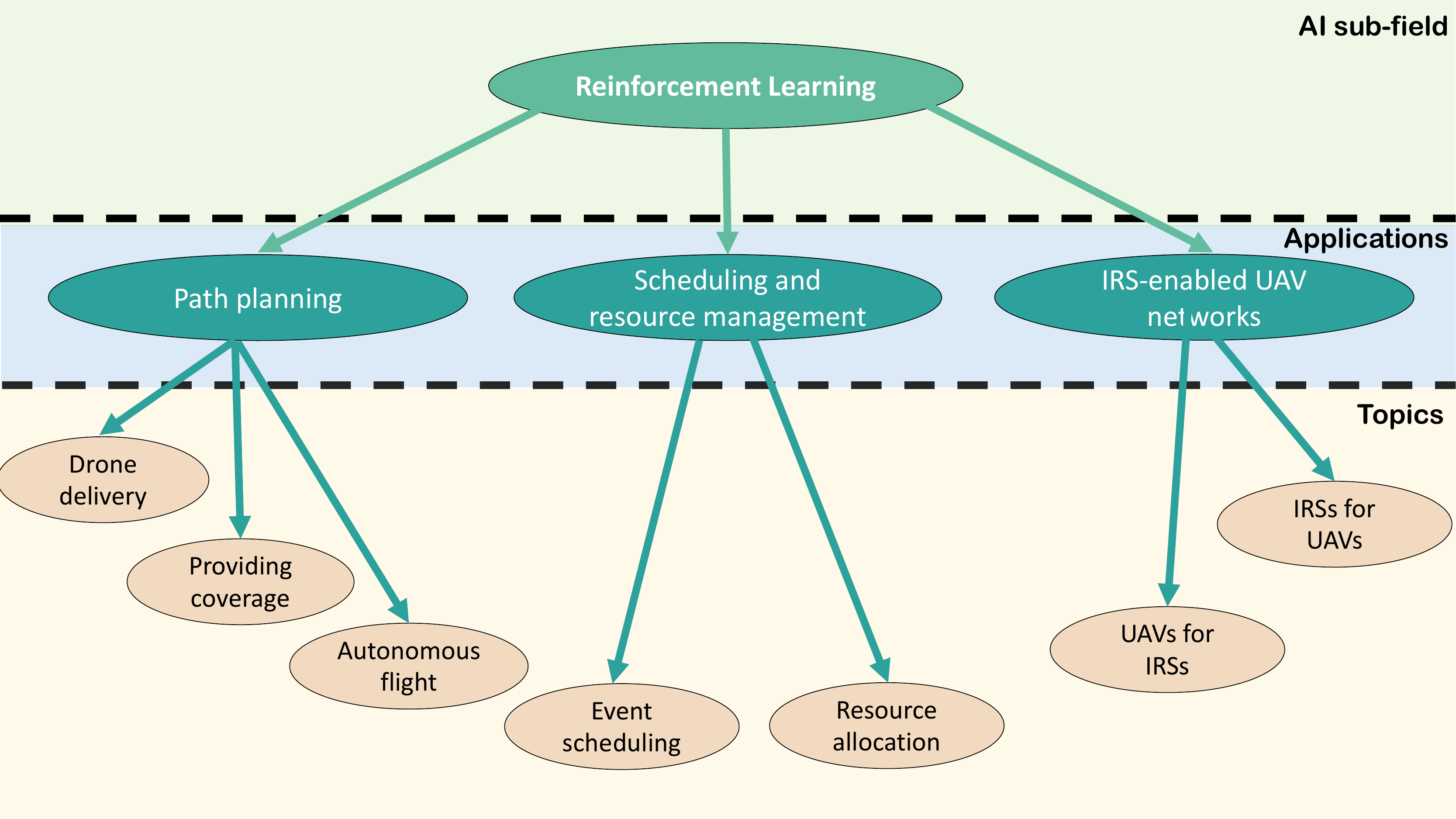}
	\caption{RL taxonomy.}
	\label{fig:taxonomy}
\end{figure*}
\subsubsection{RL limitations} 
Unlike supervised learning, RL is the area of ML that does not use the power of data to learn a task. Instead, RL uses the so-called \say{trial and error} methodology based on an agent\textquotesingle s past experiences. Surprisingly, this fact makes RL an extremely potent tool, especially for UAV-based problems such as path planning, resource management and scheduling, where data is sometimes elusive. On the other hand, RL might resemble to supervised learning in one single point which is the goal of achieving full autonomy within a UAV network by equipping UAVs with the ability to autonomously make decisions in real-time situation. Generally speaking, RL has proven its efficiency by excelling in many tasks and games, such as beating the world's top chess grandmasters. However, many still doubt the applicability of RL in real-world problems, especially for autonomous flying or autonomous driving tasks. Not only is it difficult to have a perfect perception of the environment from the agent\textquotesingle s perspective, but it is also extremely difficult to perform the exploration/exploitation dilemma explained earlier in Sec.~\ref{SEC:CaseStudy}. To be more specific, let's take the example of planning a path for a UAV to reach a target in a fully autonomous control mode. Then, in order to apply RL, the UAV, which is the agent technically speaking, needs to perform exploration to discover its surroundings and learn how to react. But this is almost impossible, especially for a high-varying environment. In other words, the exploration task is somehow impeded by the complexity of the environment and the cost of a UAV crash.\par
In terms of regulations, and as mentioned in Sec.\ref{sec:SuperAndUnsuper}, many countries do not allow the use of autonomous UAVs. For example, the use of UAVs for delivery has been completely excluded in the latest FAA regulations, as the new rules require that the UAV must always be in the operator's field of sight, which is by definition in contradiction with UAV delivery applications. Hence, such rules can impede the progress done so far in RL for different UAV applications.

Interested readers are referred to ~\cite{RLconclusion1} for more information on practical challenges that may arise in the real world application of the RL, such as system delays, cases with limited samples and more. On the bright side, several large companies and research labs have been working on producing alternatives to RL such as Evolution Strategies (ES) proposed by OpenAI~\cite{RLconclusion2} . 
To sum up, even if RL is not the ultimate solution for all UAV-based problems, it should be for some of them, which is confirmed by the numerical results of many papers covered so far in this survey. In what follows, we present our observations and criticisms of the current state of the art.

\subsubsection{Literature discussions}
Based on the recent literature related to RL for UAV- related problems, we would like to offer the following observations. One can easily notice that the big majority of the published works are focusing on path planning applications for UAVs.  More specifically, we remarked that majority of papers tend to use a Q-learning approach to propose an autonomous path planning for the UAVs. Although Q-learning is a classical algorithm and an interesting way to start solving such problems, it is somewhat impeded by the need of full knowledge of the map which is not trivial in reality, especially when considering a high movement speed of UAVs. 
Added to that the fact that  Q-learning might be slow if optimality is needed. 
Consequently, a trade-off between both complexity and optimality must be carefully studied. 
To sum up, DRL techniques such as Q-learning neural networks and DDPG, are more promising in terms of path planning and should gain more interest in the future.\par
In addition, we also noted that most research contributions use a discrete approach for path planning problems. While solutions with a discrete set of actions and states is a classic approach to address RL problems, they do not reflect a real situation where actions could be infinite as in real world trajectory planning. Although solutions with a continuous state/action space are more difficult, solving them can only bring significant benefits to the area.\par 
Furthermore, we noticed that  most of the existing work are focusing on traditional centralized approaches for RL solutions which raises several challenges related to complexity and time management. That is why we strongly believe that distributed RL is an interesting technique to solve UAVs real-time application such as the distributed Q-learning algorithm. This type of RL techniques is well suited for UAV networks where multi agents are subject to collaborative decisions.\par
Besides that, other potential applications  such as resource allocation and event scheduling are not well covered by the literature which has created an unbalanced research content oriented towards path planning problems. The actual works looking into these topics are quite few, and hence future works can be directed towards applying other RL based approaches to solve these problems. 

\section{Federated learning for UAVs } \label{sec:FL}

So far, we have covered a large number of techniques that could contribute to the development of intelligent UAV networks, ranging from supervised learning to unsupervised learning to RL. However, some of the algorithms covered previously do not go hand in hand with some UAV constraints. Specifically, we highlight the limited computing capacity on-board. Consequently, we question the applicability of AI in a UAV networks in a practical situation. In response to this last question, Google has recently implemented what is called FL, envisioning a practical way to implement ML algorithms in constrained networks~\cite{google1,google2}. FL is based on executing ML algorithms in a decentralized manner without the need to download the training set to a central node (or server). It is not specifically designed for a UAV network, but for any type of network composed of a central server (e.g. a base station in our configuration) and a number of clients (e.g. UAVs, mobile users). 

\subsection{FL Principle} 
\label{sec:FDLprinciple}
\begin{figure}
	\centering
	\includegraphics[width=3.5in]{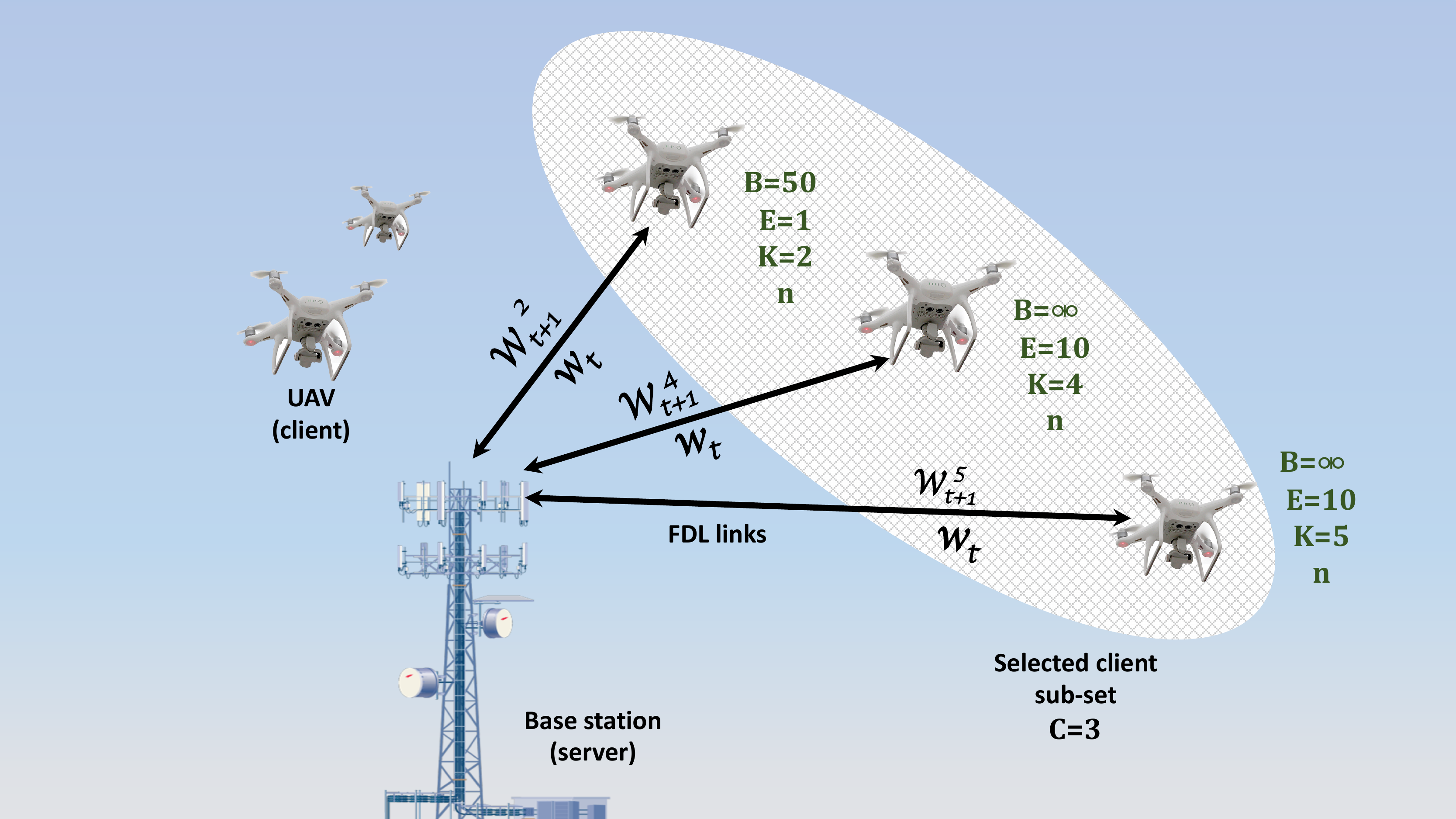}
	\caption{Federated learning principle.}
	\label{fig:FDL}
\end{figure}

Without loss of generality, we provide a comprehensive explanation for FL algorithm for a setup of a network of UAVs that are served by a terrestrial base station. As a typical task, we suppose that the UAVs are processing different ground images. We also assume that the optimization of the loss function is done through a simple stochastic gradient descent (SGD) algorithm. 
As illustrated in Fig.~\ref{fig:FDL}, the central server, which is the base station in our case, shares the current update of the global model, denoted by $w_t$, with a sub-set of the users. The subset size denoted by C, is randomly selected by the server.
Once the client UAV receives the current update of the global model, it uses its local training data to compute a local update of the global model. We should mention that several parameters are related to each UAV as indicated in Fig.~\ref{fig:FDL}. Those parameters are the mini-batch size denoted by B which indicates the amount of the local data used per each UAV, the index k of the UAV, and the the number of
training passes each client makes over its local dataset on each round, which is denoted by E. After performing the update, the UAV only communicates the update, denoted by $w_{t+1}^k$, to the base station. For an SGD based optimization, the update is calculated as follows : 
\begin{gather}
\label{eq:update}
w_{t+1}^k=w_t-\eta   \nabla l(w_t,B),
\end{gather}
where $\eta$ is the learning rate and $l$ is the loss function.\par For example, the UAV ($k=4$) on Fig.~\ref{fig:FDL} performs a full batch update and hence uses all its local data since $B=\infty$. Then it repeats the~\ref{eq:update} ten times since $E=10$ and delivers the output $w_{t+1}^k$ to the base station. 
Once the local update $w_{t+1}^k$ is received by the base station, it improves the global model and then removes these updates because they are no longer needed.   
\subsection{FL Advantages}
At the ML discussion section, we have already mentioned FL as a promising solution for constrained networks where exhaustive calculation could not be done on board. It allows to decouple the model training and the access to the raw data due to the fact that UAVs are not required to share any data with the server, instead, they only transmit their local update as explained previously.
First and foremost, FL reduces privacy and security issues by minimizing data traffic over the network. As a result, it is considered a key solution for confidential systems where data does not need to be shared. We may consider a recommender system as an example of ML application where it is recommended that raw data will not be shared between the clients. In fact, some of the clients does not want others to reveal their preferences, therefore FL preserve this privacy by keeping the local data of each user private and sharing model updates only.  
Second, FL is suitable for applications where data is unbalanced.  In other words, one client may be outside the region of interest and therefore have a small amount of data compared to other clients. Let's take the example of detecting a car using a UAV's camera, so even if one of the UAVs is misplaced in a given location where cars don't often pass by, that UAV will effectively detect a car when it is in the field of its camera. This is because the other UAVs communicating with the server have participated in the training of the misplaced drone.
In addition, the learning process within the FL can be active even if one of the nodes is in the idle state, i.e., even if one of the UAVs has to perform charging, an emergency landing or encounters a connectivity failure, the learning process continues and the UAV can restore the updates when it reconnects to the network. Lastly, FL performs well on non-independent and identical distributed data, for example,  the data partition observed by a single drone  cannot be representative of the overall data of the system simply because the UAV can only visualize part of a given process.

\subsection{FL Solutions for UAVs}
\subsubsection{FL for resource allocation and scheduling}

In~\cite{FDL6}, an optimization problem is formulated to design joint power allocation and scheduling for a UAV swarm network. The considered network is composed by one leader UAV and a group of following UAVs. Each one of the following UAVs runs a FL algorithm on its local data and then sends the update to the leader drone. The leader aggregates all the local updates in order to perform  a global update to the global model. While exchanging the updates between the UAVs, several natural phenomena will affect the wireless transmissions such as fading, wind, the transmission delay, the antenna gain and deviation, and interference. In the same mentioned work, the impact of these wireless transmission factors on the performance of the FL algorithm is analyzed. Finally, the effect of multiple wireless parameters on the convergence of the FL is studied. \par
Some of the work devoted to vehicle networks can be considered as a source of inspiration for UAV networks because there is a remarkable similarity between these two types of networks. But that's not all, the results of UAV networks are more promising because of the higher probability of establishing LOS links. In this context, the authors in~\cite{FDL1,FDL2} investigate enabling ultra-reliable low latency communication for a network of vehicular users. In short, the authors propose a joint power and resource allocation framework for ultra-reliable and low-latency network of vehicles based of FL. Surprisingly, the FL based method achieves the same performance with a centralized method but with a significant reduction of data transfer that reaches 79\%. This fact shows how FL can enable privacy in the network with the same performance as classical algorithms. \par 
In the same context, authors in~\cite{FDL10} have investigated task and resource allocation for high altitude balloon (HAB) networks. It is worth mentioning that this type of network has a lot of similarity with a UAVs network as the high-altitude balloons  are operating as a wireless base station. The authors formulate an optimization problem for a mobile edge computing enabled balloon network by minimizing energy and time consumption. However, to solve this problem, the user association with the balloons need to be specified first. To solve the latter issue, a distributed SVM-based FL model was proposed to determine to which HAB every user connects. As usual, FL principal will guarantee privacy by minimizing data sharing across the network.\par
To improve the efficiency of an internet of vehicles network, the authors in~\cite{FDL11} suggest the use of UAVs as relays in order to overcome the problem of communication failure while executing a FL task. To do so, the authors propose the formation of a coalition of UAVs in order to facilitate the training process by improving the communication efficiency level.
Each UAV in a coalition will participate in the training in a sequential way, in other words, after completing the maximum number of iterations by the nearest UAV, the second nearest UAV takes over and continues training the model and so on until all required iterations are done. A reward will be received to the UAV depending on the number of iterations performed. In the same context, an auction was designed for the  UAVs to find the optimal allocation that maximizes the profits of the drones.
So far, we covered the works that use FL to enable resource management within a UAV network, in what follows, we go over FL solutions for autonomous UAV navigation problems.

\subsubsection{FL for UAV path control}
The authors in~\cite{FDL3} investigate the control of a massive number of UAVs starting from a source point and aiming to reach a destination point. The motion of the drones is perturbed by the wind which is the main source of randomness in the problem. This disturbance can lead to a fatal collision between the UAVs and, as a solution, trajectory control is proposed to avoid this scenario. The authors used the mean field game framework to control the UAV path. However, within this framework, complex differential equations must be solved analytically, which is not possible for real-time applications and, in particular, for constrained networks.  That is why approximated solutions are proposed based on two ANNs for each one of the two differential equations. At this level, even approximating the solution via DL is not enough for the convergence of the mean field game framework. Thus, FL is used as a solution to share model parameters of the two NN between UAVs and, as a result, UAVs will be able to take into account the effect of locally non-observable samples by a UAV for learning. 

\subsubsection{FL for Flying Ad-Hoc Network (FANET) security}
FANET is a decentralized communication architecture composed of a group of UAVs where one of them at least is connected to a ground base stations or a satellite. In recent years, a significant number of research works investigated the performance and the security level of such a setup. For instance, in~\cite{FDL4}, FANET security is studied in depth. This type of network is vulnerable to jamming attacks disrupting the communication at the receiver. To avoid such scenario, the authors propose a FL assisted solution for jamming attack detection.
Many reasons stand behind selecting FL as a potential solution for FANET. First, FANETs are usually heterogeneous networks in terms of power consumption constraints and communication range. Secondly, the data available at the different nodes is unbalanced and, lastly, because the number of interacting nodes is huge. As we have already mentioned previously, FL performs well on this type of setup. 
Moreover, the FL technique is enhanced by a client selection algorithm based on Dempster-Shafer algorithm. This technique enables user group prioritization mechanism allowing selecting better clients for calculating the global update to the model.
The numerical results, based on two different datasets, show that FL always outperforms distributed learning in many different setups. Furthermore,  due to the different values of latency and bandwidth available at each UAV, the client selection based FL model itself outperformed the traditional FL algorithm. In another area of research, the authors of~\cite{FDL19} have proposed a defense strategy to deal with jamming attacks on FANETs. The jamming detection strategy is based on federated RL, more precisely, on the Q-learning algorithm. In view of the numerical results, the authors proved that the proposed defense architecture, which combines RL and FL without a model, outperformed the distributed approach.
\subsubsection{FL for content caching}
To address one of the 5G drawbacks, which is the increased delay caused by the significant activity and congestion at the backhaul links, the 6G networks employ content caching at the small-cell base stations of the 6G network. Those small-cells can be in some scenarios UAVs acting as base stations. As a result, content caching is considered as a promising alternative regarding the limited capacity of the UAVs in terms of computing capacity and memory. In this context, the work in~\cite{FDL5} investigates an intelligent caching technique for a 6G  heterogeneous aerial-terrestrial network composed of heterogeneous base stations such as UAVs and terrestrial remote radio heads. The proposed solution is based on FL techniques and hence users are no longer required to share explicitly their reporting and content preference. Instead, a heterogeneous computing platform (HCP), will accurately predict the content cached to the different base stations depending on mobile users' preferences.
In the above-mentioned setup, the HCP plays the role of the server and the different nodes of the network will only share updates to the global model in a secure manner.
A CNN was used so that the HCP learns the most popular files to cache in the heterogeneous base stations, and the optimization of the loss function is done via SGD as described previously in Sec.~\ref{sec:FDLprinciple}.
The HCP based FL solution was tested on different data-sets and proved its efficiency compared to other baseline methods. 

\subsubsection{FL for UAV sensing}
With the rapid development of the UAV industry, remote sensing has made its appearance in many different applications. The authors of~\cite{FDL16} studied the use of FL in the detection of fresh air quality by measuring the Air Quality Index (AQI). The authors used deep learning to perform the task based on images taken by the UAV network. In the same work, the authors compared the application of FL in parallel with the deep neural network instead of using other centralized methods such as SVM or CNN. The accuracy of the proposed framework was evaluated on a real data set and was found to be higher than conventional approaches.

\subsubsection{Client Selection Strategies for UAVs}

Many published works related to FL are made on an optimistic assumption that all the client will unconditionally participate in FL whenever they are called by the server. However, it is obvious that deciding which nodes (UAVs) should participate in the learning is a sensitive task for FL that could influence  the overall accuracy. Thus, in this section, we will cover some client selection technique and some participation strategies that could be of great importance to FL.\par
Authors in~\cite{FDL8} propose a contract-matching solution based on which the UAV will get a reward according to its type. The contract proposed by the authors is multi-dimensional so that it takes into account the different sources of heterogeneity in the UAV types. After setting the contracts, a matching-based algorithm will assign the optimal UAVs to each region. The UAV parameters considered while designing the contracts are the sensing model, computation model, and the transmission model.  The proposed method enabled selecting the UAVs with the lowest costs to the target sub-region. \par 
\subsection{Discussions and Future works}

We would like to emphasize the fact that FL is not necessarily applied only for UAV or mobile users networks, instead, it is being used successfully in many daily applications. For example, Google's Gboard implements the FL to learn a RNN to predict your next word when you start typing on the keyboard. However, we would like to point out that it is not clear how to select certain parameters in the FL algorithm as defined by Google in~\cite{google1}. For example, the client selection process has been defined as random, which raises the question of whether there is a better way to assign clients in each round of FL algorithm.
This last issue needs to be studied in depth for UAV networks where several parameters could affect the client selection process. From a wireless communication perspective,  channel quality, LOS condition, available data, and  battery state are crucial factors that could significantly affect the client selection process. To be specific, those parameters could make a subset of users more suitable to be selected for the FL training.\par
In addition, and in the meantime, while a large part of the scientific community asserts that the primary purpose of FL is data confidentiality, others doubt this assumption and argue that even sharing only updates over the wireless network is not safe. Unfortunately, the FL could be subject to a poisoning attack threatening the integrity of the model. These types of attacks are known in the ML community by backdoor attacks, and are typically carried out either by a single node or by a group of nodes injecting poisoned data into the model to adversely influence it.  More importantly, even FL remains vulnerable to this category of attacks not by poisoning the data but by poisoning the model itself by some malicious clients~\cite{other2,other3}. In the same vein, as a futuristic solution to the unreliability of FL systems, we propose to support UAV networks using Blockchain techniques to increase the integrity of local models at each drone. The fusion of Blockchain and FL is considered a hot topic and a number of recent works have started to study this field~\cite{blockchain1,blockchain2,blockchain3,blockchain4,blockchain5,blockchain6}. It has been found that, in addition to the increased level of stability and integrity, the Blockchain technique can increase the users motivation to participate in training by accurately rewarding them for their contribution, perhaps using cryptographic currency. Even for UAV-based networks, the research community has recently begun to apply the concept of a Blockchain coupled to FL to propose solutions for UAV networks. For example, the authors in~\cite{blockchain7} proposed a secure FL framework for a mobile crowdsensing application assisted by a UAV-network. The local exchanges of the FL algorithm were secured on the basis of a Blockchain architecture. In short, we highlight the potential of  combining  Blockchain and FL in future work.  \par 
In addition to security issues, more attention should be paid to the convergence of a FL algorithm which is not always guaranteed. Convergence depends on the specific type of a problem, such as the convexity of the loss function and the number of updates performed on the model. For example, if there is a poor selection of clients where the selective nodes are not available or do not have enough data, the optimization of the overall model will fail. One can notice that this issue overlaps with the client selection problem mentioned previously, however, it not only related to client selection but also to the type of the loss function. \par
As we proposed FL as a solution to running ML on the edge at the ML discussion section, we should mention that the massive exchange of updates across the network will result in a huge communication loads in the training phase, especially for neural networks, which will induce a scalability problem for FL. Added to that the fact that UAV networks are usually characterized by a limited battery capacity and constrained bandwidth, which makes the UAVs unable to support all this communication loads. To face this issue, many researchers have been working on alternatives and solutions that could improve memory consumption and  communication efficiency by reducing the number of communication rounds~\cite{FDL17,FDL12,FDL13,FDL14,FDL15}.\par  Another drawback of the FL arises when operating in a heterogeneous UAV network formed by various types of UAVs, rotary or fixed-wing, with different processing capabilities and different GPUs. These differences mean that some UAVs will have fast response times while others will experience severe delays. Therefore, since the FL algorithm is expected to receive the required model updates at each communication round, these induced delays will cause a major problem by considerably slowing down convergence. In some works, network nodes with large response delays are called "stragglers". In~\cite{FDL18} a distributed computation scheme has been proposed to mitigate the impact of slow nodes on convergence for gradient methods. In addition, the quality of connectivity can affect the convergence of the FL algorithm since several network nodes may encounter an unexpected failure when transmitting their local updates. These interruptions can also degrade the overall efficiency of the FL by slowing convergence.  

To sum up, even with all the above-mentioned issues related to FL, it remains a good alternative for UAV-based networks. However, there are still some open problems that are worth investigation. For instance, the application of FL with supervised learning techniques, which were discussed in Sec.\ref{Sec:SupervisedML}, is still an open problem.  
\section{Conclusion}

Motivated by wide set of new applications that can benefit from UAV networks, such as smart cities and aerial BS deployment, in this paper, we have explored in detail a new research direction where ML techniques are used to enhance the performance of UAV networks. We started by providing an extensive overview for unsupervised and supervised ML techniques that have been applied in UAV networks. Then, we briefly introduced the RL technique and discussed a number of the relevant works that implemented this ML technique for UAVs. Finally, we discussed a set of research works where FL techniques are used in the area of UAV networks. For each of the three considered techniques, we provided a set of concluding remarks that discuss the current limitations and challenges as well as a set of interesting open problems.

\bibliographystyle{IEEEtran}
\bibliography{SubmissionVersion}
\begin{IEEEbiography}[{\includegraphics[width=1in,height=1.25in,clip,keepaspectratio]{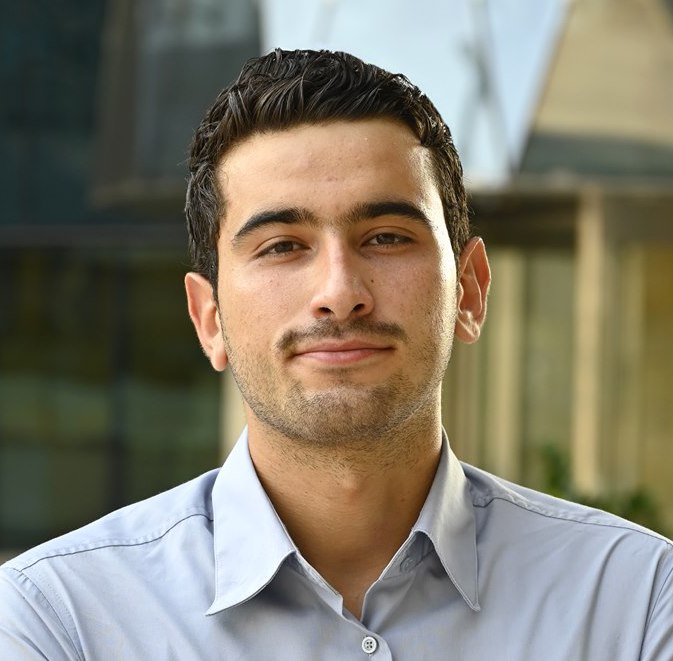}}]{Mohamed-Amine Lahmeri}(S\textquotesingle21) was born in Tunis, Tunisia. He received
	the National Engineering Diploma from Ecole Polytechnique de Tunisie.He is currently an MS/Ph.D. student with the Communication
	Theory Lab, King Abdullah University of Science
	and Technology (KAUST), Thuwal, Saudi Arabia. His current research interests include UAV-enabled communication systems, and machine learning for wireless communication.	
\end{IEEEbiography}
\begin{IEEEbiography}[{\includegraphics[width=1in,height=1.25in,clip,keepaspectratio]{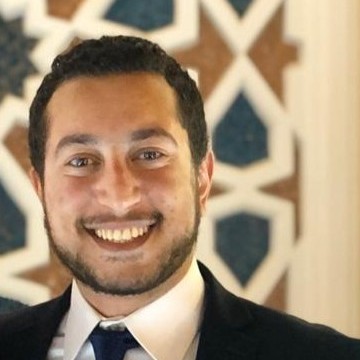}}]{Mustafa A. Kishk}(S\textquotesingle16) received
	the B.Sc. and M.Sc. degrees from Cairo University, Giza, Egypt, in 2013 and 2015, respectively,
	and the Ph.D. degree from Virginia Tech, Blacksburg, VA, USA, in 2018. He is currently a Postdoctoral Research Fellow with the Communication
	Theory Lab, King Abdullah University of Science
	and Technology (KAUST), Thuwal, Saudi Arabia. His current research interests include stochastic geometry, energy harvesting wireless networks,
	UAV-enabled communication systems, and satellite communications.	
\end{IEEEbiography}
\begin{IEEEbiography}[{\includegraphics[width=1in,height=1.25in,clip,keepaspectratio]{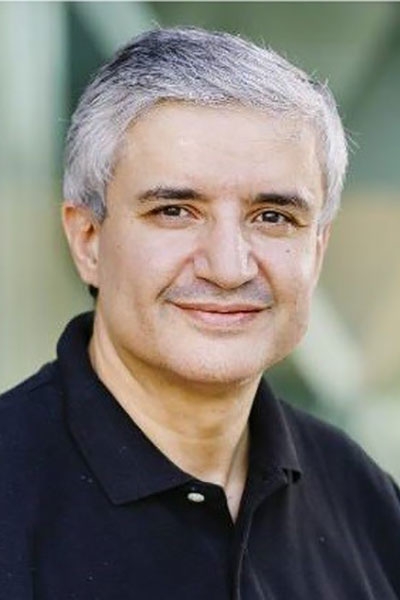}}]{Mohamed-Slim Alouini}
	(S\textquotesingle94-–M\textquotesingle98-–SM\textquotesingle03-–F\textquotesingle09) was born in Tunis, Tunisia. He received the Ph.D. degree in Electrical Engineering from the California Institute of Technology (Caltech), Pasadena, CA, USA, in 1998. He served as a faculty member in the University of Minnesota, Minneapolis,MN, USA, then in the Texas A\&M University at Qatar, Education City, Doha, Qatar before joining King Abdullah University of Science and Technology (KAUST), Thuwal, Makkah Province, Saudi Arabia as a Professor of Electrical Engineering in 2009. His current research interests include the modeling, design, and performance analysis of wireless communication systems.\end{IEEEbiography}

\end{document}